\documentclass{article}
\usepackage[margin=1in]{geometry}
\usepackage[T1]{fontenc}
\usepackage{amssymb,amsthm,mathtools}
\usepackage{multirow,makecell}
\usepackage{booktabs,float}
\usepackage[round]{natbib}

\usepackage{algorithm}
\usepackage{algpseudocode}
\usepackage{placeins}
\usepackage{xcolor}
\usepackage[hidelinks]{hyperref}
\hypersetup{
    colorlinks=true,
    linkcolor=blue,
    citecolor=blue}
\newtheorem{theorem}{Theorem}[section]
\newtheorem{proposition}[theorem]{Proposition}
\newtheorem{lemma}[theorem]{Lemma}
\theoremstyle{definition}
\newtheorem{assumption}[theorem]{Assumption}
\numberwithin{equation}{section}
\newcommand{\R}{\mathbb R}
\newcommand{\E}{\mathbb E}
\newcommand{\Var}{\operatorname{Var}}
\newcommand{\Cov}{\operatorname{Cov}}
\newcommand{\tr}{\operatorname{tr}}
\newcommand{\diag}{\operatorname{diag}}
\newcommand{\op}{\operatorname{op}}
\newcommand{\Frob}{\operatorname{F}}

\begin{document}

\title{Inference for Directions in Canonical Correlation Analysis\\ with Partially Paired Data}
\author{Chau Tran and Xiaodong Li\\
Department of Statistics and Data Science, University of California, Davis}

\maketitle

\begin{abstract}
Canonical correlation analysis (CCA) is a classical tool for learning linear associations between two data views, and inference on canonical directions is the key for interpreting their coefficients and detecting interesting relationships. We study the problem of hypothesis testing for canonical directions when paired observations are supplemented by unpaired observations of either view. The key challenge is to incorporate the additional marginal information while accounting for its dependence on the paired sample used to estimate the canonical directions. To address this issue, we propose coupled augmentation functions that jointly use the marginal information from both views and construct a one-step estimator based on the resulting corrections. Theoretically, we show that under the null, the proposed estimator is consistent and asymptotically normal even when dimension grows at the rate $d = o(n)$. This result provides a sign-invariant Wald test statistic for detecting important variables in canonical directions. Through numerical studies, we show that the proposed test statistic with semi-paired data improves upon inference using paired data.
\end{abstract}

\section{Introduction}\label{sec:introduction}

Canonical correlation analysis (CCA) identifies maximally correlated linear combinations of two sets of measurements, or data views \citep{Hotelling1936}. Canonical directions are increasingly used to characterize associations across multi-view and multi-modal data. For example, their coefficients can identify variables that contribute to cross-view relationships in biological data integration and brain-behavior association studies \citep{LeCaoEtAl2009,SmithEtAl2015, lu2024integrative}. However, CCA has traditionally been used primarily as an exploratory tool with most existing works focusing on estimation \citep{witten2009penalized, gao2017sparse, ma2020subspace, wang2020finding}. At the same time, inference is essential for interpreting CCA results and distinguishing interesting associations from potentially spurious findings \citep{dinga2019evaluating, zhuang2020technical, HelmerEtAl2024}. This motivates the development of rigorous inferential procedures for canonical weights. Specifically, we seek to determine which individual variables make statistically significant contributions to the linear associations between the two data views.

While paired data are available for most sampling units, unpaired observations are also common \citep{FerreiraEtAl2022, ShafieiEtAl2025}.Most methods use only paired observations, even though unpaired observations contain information about the marginal distributions of the two views. Although unpaired observations do not supply additional cross-products between the views, they can improve estimation of the marginal covariance structures and thereby reduce uncertainty in the canonical directions. We therefore aim to use this additional marginal information to improve inference on coefficients of canonical directions.

Formally, we observe $n$ i.i.d. paired samples $(X_i,Y_i)$, where $X_i \in \R^p$ and $Y_i \in \R^q$, together with independent unpaired samples of sizes $N_X$ and $N_Y$, either of which may be zero. Our goal is to test hypotheses of the form $H_{0j}: u_{1j}^*=0$ or $H_{0k}: v_{1k}^*=0$, where $u_{1j}^*$ and $v_{1k}^*$ denote coordinates of the leading $X$-side and $Y$-side canonical directions, respectively. Incorporating unpaired observations creates two issues. First, the adjustments that use the additional $X$-only and $Y$-only observations are coupled through the paired sample, so we need to jointly determine augmentation functions for both views. Second, these functions are estimated using the plug-in CCA estimates from paired data, which introduces higher-order error terms as the dimension grows with sample size. For valid inference, we must establish asymptotic normality under the null by showing that these remainder terms are negligible.

\paragraph{Contributions.}
First, we derive optimal augmentation functions that jointly minimize the asymptotic variance of the corrected estimator. Under joint Gaussianity, these functions have explicit forms, so the resulting one-step estimator uses CCA estimates from paired data and pooled marginal covariance estimates. Second, under regularity conditions, we show that the proposed test statistic has an asymptotic $\chi_1^2$ distribution under the null when $d=p+q=o(n)$. The Wald test statistic is invariant to the signs of CCA estimates which make it practical for inference. We also extend the approach to confidence intervals for the leading canonical correlation using an analytic bias correction when dimension grows at rate $d=o(n^{2/3})$. Through numerical studies, we demonstrate the favorable performance of the proposed one-step estimator and Wald test statistic when unpaired observations are available.

\paragraph{Inference for CCA.}
Classical theoretical works on CCA provide asymptotic distributions of estimators of canonical correlations and canonical directions when dimensions $p$ and $q$ are fixed and the sample size $n \rightarrow \infty$ \citep{Anderson1999, Anderson2003, TaskinenEtAl2006, Ogasawara2007}. In modern applications, however, it is important to allow dimensions to grow with the sample size \citep{LahaEtAl2023, bykhovskaya2023high, pu2026asymptotic} and to accommodate incompletely paired observations \citep{maleki2026high}. We develop hypothesis tests for individual coordinates of the leading canonical directions when dimensions grow with sample size. Our method accommodates additional unpaired observations from either view and, unlike the bootstrap approach of \citet{KesslerLevina2023}, does not require resampling. Previous works have also developed hypothesis tests for canonical correlations to assess linear relationships between the views \citep{WinklerRenaudSmithNichols2020, mckeague2022significance}. Although our primary focus is on inference for canonical directions, we extend our method to construct confidence intervals for the leading canonical correlation.

\paragraph{CCA with partially paired data.}
Existing methods such as SemiCCA \citep{KimuraEtAl2010} and semi-paired probabilistic CCA \citep{ZhangEtAl2014} use unpaired observations to improve CCA estimation, whereas we study hypothesis tests for canonical directions. More closely related to our work, \citet{Yamada2013} derives asymptotic distributions for Gaussian maximum likelihood estimators of canonical correlations and directions with additional observations from one view, assuming fixed dimensions. We instead develop a one-step correction that accommodates additional observations from both views and establish valid tests with growing dimensions.

\paragraph{Semi-supervised inference.}
Our approach builds on the projection principle used in semi-supervised inference \citep{ChakraborttyCai2018, ZhangBrownCai2019, SongLinZhou2024, XuWittenShojaie2025}. Under a Gaussian population assumption, we derive jointly optimal corrections for estimators from both views.

\section{Setup and background}\label{sec:method}\label{sec:main}

\subsection{Preliminaries}
Consider i.i.d. pairs $Z_i = (X_i^\top,Y_i^\top)^\top\sim N(0,\Sigma)$,
$i=1,\ldots,n$, where $X_i\in\R^p$, $Y_i\in\R^q$, and
\[
 \Sigma=\begin{pmatrix}\Sigma_X&\Sigma_{XY}\\
                         \Sigma_{YX}&\Sigma_Y\end{pmatrix}.
\]
Let $d=p+q$ and $m=\min(p,q)$. Throughout, we use the convention $\lambda_\ell=0$ for $\ell>m$. In particular, when $m=1$ the eigengap condition below is interpreted with $\lambda_2=0$. We make the following assumption on the covariance matrix.
\begin{assumption}[Uniform Gaussian CCA conditions]\label{ass:model}
For an absolute constant $c >0$,
\[
 c \le\lambda_{\min}(\Sigma)\le\lambda_{\max}(\Sigma)\le c^{-1},
 \qquad \lambda_1^2-\lambda_2^2\ge c,
\]
where $\lambda_{\min}(\Sigma)$ and $\lambda_{\max}(\Sigma)$ are minimum and maximum eigenvalues of $\Sigma$, and $\lambda_1\ge\lambda_2\ge\cdots \ge \lambda_m \ge 0$ are the singular values of $\Sigma_X^{-1/2}\Sigma_{XY}\Sigma_Y^{-1/2}$.
\end{assumption}
This assumption requires the joint and marginal covariance matrices to be well-conditioned. More importantly, it ensures a positive eigengap so that, after fixing their sign, the leading canonical directions are uniquely defined and differentiable in the neighborhood of the population covariance $\Sigma$. The leading CCA directions $u_1^*,v_1^*$ are maximizers of the objective
\begin{equation}\label{eq:cca-est}
	\begin{split}
		\max_{u \in \R^p, v \in \R^q} u^\top \Sigma_{XY} v \qquad \text{subject to} \qquad u^\top \Sigma_X u=1 \text{ and } v^\top \Sigma_Y v=1.
	\end{split}
\end{equation}
Let $\widehat\Sigma = n^{-1}\sum_{i=1}^n Z_iZ_i^\top$ be the sample covariance matrix and denote its blocks by $\widehat\Sigma_X,\widehat\Sigma_{XY},\widehat\Sigma_Y$. The sample canonical triplets are $(\widehat u_\ell,\widehat v_\ell,\widehat\lambda_\ell)$, $\ell\le m$. Note that the CCA estimates are normalized so that for $1 \le \ell, k \le m$
\[
 \widehat u_\ell^\top\widehat\Sigma_X\widehat u_k
 =\widehat v_\ell^\top\widehat\Sigma_Y\widehat v_k
 =\mathbf1\{\ell=k\}.
\]
For notational convenience, when $p>m$, we extend $u_1^*,\ldots,u_m^*$ to a $\Sigma_X$-orthonormal basis of \(\mathbb R^p\) and set \(\lambda_\ell=0\) for \(\ell>m\). We define the sample quantities analogously.

Optionally, we observe i.i.d unpaired samples $X_r^U\sim N(0,\Sigma_X)$, $r = 1, \dots, N_X$, and $Y_r^U\sim N(0,\Sigma_Y)$, $r = 1, \dots, N_Y$, with the two unpaired samples mutually independent and independent of the paired samples. The additional observations have the same marginal distributions as paired samples, which is a common assumption in semi-supervised settings \citep{XuWittenShojaie2025, GrahamCaroneRotnitzky2026}. Define the unpaired fractions in the pooled samples
\begin{equation}\label{eq:gammas}
 \gamma_X=\frac{N_X}{n+N_X},\qquad
 \gamma_Y=\frac{N_Y}{n+N_Y}.
\end{equation}
For $X$, we can define estimates of marginal covariance matrices using the pooled data as
\begin{equation}\label{eq:pooled-marginals}
 \widehat\Sigma_{X,\mathrm{pooled}}
 =\frac{\sum_{i=1}^nX_iX_i^\top+
          \sum_{r=1}^{N_X}X_r^U(X_r^U)^\top}{n+N_X},
\end{equation}
and write $D_X=\widehat\Sigma_{X,\mathrm{pooled}}-\widehat\Sigma_X$. If $N_X = 0$, then $D_X = 0$ and this estimate coincides with the estimate from paired data. $D_Y$ is defined similarly. Our goal is to leverage the pooled estimates to improve estimation and inference when additional observations are available.

\subsection{Optimal augmentations with semi-paired data}\label{sec:opt-aug}
To gain the intuition behind our methods, we first consider paired observations. Let $g_j(X,Y)$ denote the influence function of $u_{1j}^*$ (see Appendix~\ref{sec:augmentation-derivation} for more details). By Taylor expansion
\begin{equation}\label{eq:gen-eq}
\widehat u_{1j}-u_{1j}^*
=\frac{1}{n}\sum_{i=1}^n g_j(X_i,Y_i)+\text{remainder terms}.
\end{equation}
When the remainder is negligible, inference can be based on the limiting distribution of $\frac{1}{n}\sum_{i=1}^n g_j(X_i,Y_i)$ \citep{TaskinenEtAl2006, Ogasawara2007}, which we will discuss in Section~\ref{sec:theoretical}. Next, suppose that unpaired observations are available only for $X$. For a function $a_j$ of $X$, we can write $g_j(X,Y)=\{g_j(X,Y)-a_j(X)\}+a_j(X)$. The first term is estimated using paired observations, whereas the second term can be estimated using the pooled $X$ observations. We therefore incorporate the additional observations by replacing the paired-sample mean of $a_j(X)$ with its pooled-sample mean, as in semi-supervised inference literature \citep{ZhangBrownCai2019, SongLinZhou2024, XuWittenShojaie2025}. Ideally, we want $a_j(X)$ to capture the variation of $g_j(X,Y)$ that can be learned from $X$ alone. The optimal choice is
\begin{equation*}
	a_j^*(X)=\E\{g_j(X,Y)\mid X\}.
\end{equation*}
When unpaired observations from both views are available, we make a corresponding adjustment using a function $b_j(Y)$. The two adjustments share the paired sample, so their covariance must be taken into account. We therefore choose $a_j$ and $b_j$ jointly to minimize the variance
\begin{equation}\label{eq:main-augmentation-variance}
	\begin{split}
		V_j(a_j,b_j) = &\Var[g_j(X, Y) - \gamma_X a_j(X) - \gamma_Y b_j(Y)] \\
		&+ \gamma_X(1-\gamma_X)\Var[a_j(X)] + \gamma_Y(1-\gamma_Y)\Var[b_j(Y)].
	\end{split}
\end{equation}
The first term comes from the paired sample, and the remaining terms come from the independent unpaired samples. We formalize the optimal augmentations by the following proposition.

\begin{proposition}[Optimal augmentations]
\label{prop:main-projection}
Let $P_{XY}$ denote the joint distribution of $(X,Y)$, with marginal distributions $P_X$ and $P_Y$. Denote the space of functions with mean zero and finite second moment as
\[L_0^2(P) = \left\{f\in L^2(P):\int f\,dP=0\right\}.\]
Suppose that $g_j\in L_0^2(P_{XY})$ and $0 < \gamma_X,\gamma_Y<1$. Then, the functional $V_j$ defined in \eqref{eq:main-augmentation-variance} has a unique minimizer
\[
(a_j^*,b_j^*)
\in L_0^2(P_X)\times L_0^2(P_Y)
\]
that is characterized by
\begin{equation}\label{eq:main-projection-equations}
\begin{aligned}
a_j^*(X)
+\gamma_Y\E\{b_j^*(Y)\mid X\}
&=\E\{g_j(X,Y)\mid X\},\\
b_j^*(Y)
+\gamma_X\E\{a_j^*(X)\mid Y\}
&=\E\{g_j(X,Y)\mid Y\}.
\end{aligned}
\end{equation}
\end{proposition}
Appendix~\ref{sec:generic-projection} provides the derivation and discusses the cases where either unpaired sample is absent.

\section{Methods}
\label{sec:main-projection}
\subsection{One-step estimator}\label{sec:coordinate-method}
Under joint Gaussianity, the optimal augmentation functions in Proposition~\ref{prop:main-projection} have explicit forms. We denote them by $a_j(X)$ and $b_j(Y)$ throughout the remainder of the paper, and their expressions and derivation are given in Appendix~\ref{sec:explicit-augmentations}. We use hats to denote plug-in estimates of population quantities. Then, the one-step estimator is
\begin{equation}\label{eq:ss-augmentation-estimator}
\begin{aligned}
\widehat u_{1j}^{\mathrm{Semi}} ={}& e_j^\top\widehat u_1 +\overline{\widehat a}_{j,\mathrm{pooled}} -\overline{\widehat a}_{j,\mathrm{paired}} +\overline{\widehat b}_{j,\mathrm{pooled}} -\overline{\widehat b}_{j,\mathrm{paired}}.
\end{aligned}
\end{equation}
Here, the bars denote averages over the pooled or paired samples, with both averages evaluated using the same paired CCA estimates. Thus, each adjustment replaces a paired-sample average with a pooled-sample average. Since the augmentation functions are linear combinations of products of canonical variates, their sample averages depend on the unpaired observations only through the pooled second-moment matrices. Therefore, the unpaired data enter the estimator through the covariance differences $D_X$ and $D_Y$. The estimator can therefore be computed as
\begin{equation}\label{eq:ss-estimator}
\widehat u_{1j}^{\mathrm{Semi}}
=
e_j^\top\widehat u_1
+
\sum_{\ell=1}^m
(e_j^\top\widehat u_\ell)\widehat h_\ell.
\end{equation}
The corrections $\widehat h_\ell$ are shared across all coordinates. For $2\le\ell\le m$, write
\[
\widehat{\mathcal D}_{\ell,\gamma}
=(\widehat\lambda_1^2-\widehat\lambda_\ell^2)
 (1-\gamma_X\gamma_Y\widehat\lambda_1^2\widehat\lambda_\ell^2).
\]
The shared corrections are
\begin{equation}\label{eq:shared-coordinate-correction}
\begin{aligned}
\widehat h_1
={}&
-\frac{1-\gamma_Y\widehat\lambda_1^4}
       {2(1-\gamma_X\gamma_Y\widehat\lambda_1^4)}
 \widehat u_1^\top D_X\widehat u_1 -\frac{\widehat\lambda_1^2(1-\gamma_X)}
        {2(1-\gamma_X\gamma_Y\widehat\lambda_1^4)}
 \widehat v_1^\top D_Y\widehat v_1,\\
\widehat h_\ell
={}&
\frac{1-\widehat\lambda_1^2}
     {\widehat{\mathcal D}_{\ell,\gamma}}
\Bigl[
 \widehat\lambda_\ell^2
 (1-\gamma_Y\widehat\lambda_1^2)
 \widehat u_1^\top D_X\widehat u_\ell +
 \widehat\lambda_1\widehat\lambda_\ell
 (1-\gamma_X\widehat\lambda_\ell^2)
 \widehat v_1^\top D_Y\widehat v_\ell
\Bigr],
\end{aligned}
\end{equation}
for $2\le\ell\le m$. Without unpaired observations ($D_X=D_Y=0$), this coincides with the paired CCA estimator. The estimator for $v_{1k}^*$ follows by swapping $(X,p,u,\gamma_X)$ with $(Y,q,v,\gamma_Y)$.

\subsection{Asymptotic variance and test statistic}
In this section, we derive the asymptotic variance under the null to obtain the studentized test statistic. For $2\le \ell\le p$, the variance of the $\ell$-th influence-function component using paired data only is
\[
\tau_{0,\ell}^2 = \frac{(1-\lambda_1^2) \bigl(\lambda_1^2+\lambda_\ell^2 -2\lambda_1^2\lambda_\ell^2\bigr)}{(\lambda_1^2-\lambda_\ell^2)^2}
\]
and the variance reduction by using unpaired observations is
\begin{equation}\label{eq:modal-gain-main}
\begin{aligned}
\Delta_{\gamma,\ell} ={}& \frac{(1-\lambda_1^2)^2} {(\lambda_1^2-\lambda_\ell^2)^2(1-\gamma_X\gamma_Y\lambda_1^2\lambda_\ell^2)}
\times
\Bigl\{
\gamma_X\lambda_\ell^4(1-\gamma_Y\lambda_1^2)
+
\gamma_Y\lambda_1^2\lambda_\ell^2
(1-\gamma_X\lambda_\ell^2)
\Bigr\}.
\end{aligned}
\end{equation}
Let $\tau_{\gamma,\ell}^2=\tau_{0,\ell}^2-\Delta_{\gamma,\ell}$. Then the asymptotic variance for inference under the null is
\begin{equation}\label{eq:sigma-two-sided}
\sigma_{j,n}^2 = \sum_{\ell=2}^p (e_j^\top u_\ell^*)^2 \tau_{\gamma,\ell}^2.
\end{equation}
Substituting the sample estimates of the canonical triplets gives
\begin{equation}\label{eq:variance-estimator}
 \widehat\sigma_j^2=\sum_{\ell=2}^p(e_j^\top\widehat u_\ell)^2\widehat\tau_{\gamma,\ell}^2.
\end{equation}
Then, the test statistic under the null is given by
\begin{equation}\label{eq:coordinate-wald}
 W_j=\frac{n(\widehat u_{1j}^{\mathrm{Semi}})^2}{\widehat\sigma_j^2}.
\end{equation}

\begin{algorithm}[!t]
\caption{Test statistics for $u_{1j}$}\label{alg:ss-cca}
\label{sec:implementation-workflow}
\begin{algorithmic}[1]
\Require $n$ paired observations, $N_X$ $X$-only and $N_Y$ $Y$-only observations, target coordinates.
\State Compute $\widehat\Sigma$ and obtain CCA estimates $(\widehat u_\ell,\widehat v_\ell,\widehat\lambda_\ell)$, $1 \le \ell\le m$ using paired data.
\State Compute the pooled marginal covariances in \eqref{eq:pooled-marginals} and the differences $D_X,D_Y$.
\State Compute the corrections $\widehat h_\ell$ in \eqref{eq:shared-coordinate-correction}.
\For{each target coordinate $j$}
 \State Compute $\widehat u_{1j}^{\mathrm{Semi}}$ using \eqref{eq:ss-estimator} and $\widehat\sigma_j^2$ using \eqref{eq:variance-estimator}.
 \State Compute $W_j$ in \eqref{eq:coordinate-wald}.
\EndFor
\State \Return Test statistics $W_j$ for each target coordinate $j$.
\end{algorithmic}
\end{algorithm}

Algorithm~\ref{alg:ss-cca} summarizes the computation of the test statistic in \eqref{eq:coordinate-wald}. For dense matrices, the cost is $O(nd^2+N_Xp^2+N_Yq^2+d^3)$, with $O(d^2)$ memory when covariance matrices are stored.

\subsection{Asymptotic analysis}
\label{sec:theoretical}

\begin{theorem}[Limiting distribution]\label{thm:main}
Suppose that Assumption~\ref{ass:model} holds, and $d=o(n)$. Let $j\in\{1,\ldots,p\}$ be a coordinate satisfying $u_{1j}^*=0$. Then there exist constants $0<c_0<C_0<\infty$, such that
\[
 c_0 \le \sigma_{j,n}^2(\gamma_X,\gamma_Y) \le C_0
\]
for all $(\gamma_X,\gamma_Y)\in[0,1]^2$.

Moreover, for any $N_X,N_Y\ge0$, the variance estimator in \eqref{eq:variance-estimator} satisfies
\[
\begin{gathered}
 \widehat\sigma_j^2/\sigma_{j,n}^2\xrightarrow{p}1,
\end{gathered}
\]
and the test statistic in \eqref{eq:coordinate-wald} satisfies
\[
W_j \rightsquigarrow\chi_1^2.
\]
The test that rejects the null hypothesis $H_0: u_{1j}^* = 0$ when $W_j>\chi^2_{1,1-\alpha}$ has asymptotic size $\alpha$ for any $\alpha\in(0,1)$. The corresponding result for a $Y$-side coordinate follows by exchanging the two views.
\end{theorem}

Theorem~\ref{thm:main} establishes the limiting distribution of the test statistic in \eqref{eq:coordinate-wald} under the null. The bounds on the asymptotic variance ensure that it remains nondegenerate and well-behaved. The $p$-value is $2\{1-\Phi(\sqrt{W_j})\}$, where $\Phi$ is the standard normal distribution function. Notice that $W_j$ is invariant to the signs of canonical directions, making it practical for inference on CCA results.

To understand the condition on growing dimension, notice that we use plug-in estimates $\widehat u_{1j}$ for paired data and $\widehat u_{1j}^{\mathrm{Semi}}$ for semi-paired data. The plug-in estimates are known to face a barrier in the dimension $d$ and may fail to be $\sqrt{n}$-consistent and asymptotically normal unless $d = o(n^{1/2})$ \citep{portnoy1988asymptotic, lin2024worthwhile}. For example, bounding the remainder terms in \eqref{eq:gen-eq} by $\|\widehat\Sigma-\Sigma\|_{\mathrm{op}}^2$ will result in an $O_p(d/n)$ bound, which is $o_p(n^{-1/2})$ when $d=o(n^{1/2})$. In this paper, we utilize properties of the Gaussian distribution to obtain a sharper remainder bound under the null $u_{1j}^*=0$
\begin{equation}\label{eq:coordinate-remainder-main}
 \widetilde u_{1j}^{\mathrm{Semi}}-u_{1j}^*
 =\mathcal L_{j,n}+O_p\left(\frac{\sqrt d}{n}\right),
\end{equation}
where $\mathcal L_{j,n}$ is first-order term defined in Appendix~\ref{sec:coordinate-theorem-proof}. The remainder is $o_p(n^{-1/2})$ when $d=o(n)$, so this first-order term determines the limiting distribution as mentioned above. Here $\widetilde u_{1j}^{\mathrm{Semi}}$ denotes the estimator after aligning its sign with the population canonical direction. Its square equals $(\widehat u_{1j}^{\mathrm{Semi}})^2$, so computing $W_j$ requires no information on the orientations of the population canonical directions.

\subsection{When do unpaired observations help?}\label{sec:canonical}
Since we did not make any assumption on $N_X$ and $N_Y$, a natural question is when the unpaired observations are beneficial. We characterize this in the following proposition.

\begin{proposition}[Variance reduction]\label{prop:var-reduction}
Suppose Assumption~\ref{ass:model} holds and $u_{1j}^*=0$. Let $\sigma_{j,n}^2(\gamma_X,\gamma_Y)$ be defined as in \eqref{eq:sigma-two-sided}. Then
\[\sigma_{j,n}^2(\gamma_X,\gamma_Y) < \sigma_{j,n}^2(0,0)
\]
if and only if $\gamma_X+\gamma_Y>0$ and \[
\exists\,\ell\in\{2,\ldots,m\}:
 (e_j^\top u_\ell^*)\lambda_\ell\ne0.
\]
\end{proposition}
Proposition~\ref{prop:var-reduction} states that for a coordinate satisfying $u_{1j}^*=0$, unpaired observations reduce the asymptotic variance if and only if that coordinate has a nonzero coefficient in some non-leading canonical direction with nonzero canonical correlation.

\subsection{Confidence interval for the leading canonical correlation}
\label{sec:lambda-extension}
While our main focus is inference for the leading canonical directions, we can also extend the approach to construct confidence intervals for the leading canonical correlation $\lambda_1$, which measures the strength of association between the two views. In this case, a debiasing step is needed, and the resulting semi-supervised estimator $\widehat\lambda_1^{\mathrm{Semi}}$ is given in \eqref{eq:lambda-ss-estimator}. Let $\sigma_{\lambda,n}^2$ denote its asymptotic variance and $\widehat\sigma_\lambda^2$ the corresponding plug-in variance estimator. We have the following result.
\begin{theorem}[Leading canonical correlation]\label{thm:lambda-ci}
Suppose that Assumption~\ref{ass:model} holds and $d=o(n^{2/3})$. For any $N_X,N_Y\ge0$,
\begin{equation*}
	\widehat\sigma_\lambda^2/\sigma_{\lambda,n}^2\xrightarrow{p}1
\end{equation*}
and
\[
 \frac{\sqrt n(\widehat\lambda_1^{\mathrm{Semi}}-\lambda_1)}
      {\widehat\sigma_\lambda}\rightsquigarrow N(0,1).
\]
\end{theorem}

We then can use the distributional statement in Theorem~\ref{thm:lambda-ci} to construct confidence interval for $\lambda_1$ with asymptotic coverage guarantee. Appendix~\ref{sec:lambda-method} gives details on the estimator, bias correction step, and proof of Theorem~\ref{thm:lambda-ci}. The dimensional condition, $d=o(n^{2/3})$, is similar to that considered by \citet{lin2024worthwhile}, where the authors use jackknife correction for valid inference for functionals estimated using high-dimensional $Z$-estimators.

\section{Experiments}\label{sec:results}\label{sec:simulation-setup}
We evaluate performance using empirical size and power, reported as percentages. When $N_X = N_Y = 0$, we denote the test statistic as Paired-Wald (PW). When additional unpaired observations are used, we denote the test statistic as Semi-Wald (SW). We also compare our proposed methods with a computational bootstrap method proposed by \citet{KesslerLevina2023}, which we denotes CB. CB uses $B=2000$ bootstrap resamples with paired data, and align the signs of estimates using the weighted Hungarian algorithm. The bootstrap $p$-value is given by $\widehat p_j = 2\min\{\widehat Q_j,1-\widehat Q_j\}$, where $\widehat Q_j=B^{-1}\sum_{b=1}^{B}\mathbf1\left\{\widehat\theta_j^{(b)}\le0\right\}$ and $\widehat\theta_j^{(b)}$ is the $b$-th CB estimate of $j$-th coordinate after signs alignment for $1 \le b \le B$. For all experiments, we use significance level $\alpha=0.05$, and results are averaged over 500 replications.

\subsection{Simulated data with a rank-two model}\label{sec:rank-two-comparison}
We first consider the setting where there are two non-zero canonical correlations. Let $p=q$ be even, set $s=p/2$ to be the number of non-zero entries, and define the marginal covariance matrices by
\[
 \Sigma_X=\Sigma_Y=(0.3^{|j-k|})_{j,k=1}^p.
\]
Let $w_p=(0.2,\mathbf1_{s-1}^{\top},\mathbf0_{p-s}^{\top})^{\top}, z_p=e_1+e_p$.
The first two $X$-side canonical directions are
\begin{align*}
 u_1^*=\frac{w_p}{(w_p^\top\Sigma_Xw_p)^{1/2}}, \qquad u_2^*=\frac{\widetilde u_2}{(\widetilde u_2^\top\Sigma_X\widetilde u_2)^{1/2}}
\end{align*}
where $\widetilde u_2=z_p-(u_1^{*\top}\Sigma_Xz_p)u_1^*$. The $Y$-side directions are constructed analogously. The cross-covariance matrix is given by
\begin{equation*}
 \Sigma_{XY}=\Sigma_X[\lambda_1u_1^*(v_1^*)^\top
 +\lambda_2u_2^*(v_2^*)^\top]\Sigma_Y,
\end{equation*}
where $(\lambda_1,\lambda_2)=(0.85,0.75)$. We then sample $n$ i.i.d pairs $(X_i, Y_i)$ from the multivariate Gaussian distribution $N_d(0, \Sigma)$, together with $N_X$ and $N_Y$ i.i.d observations from $N_p(0, \Sigma_X)$ and $N_q(0, \Sigma_Y)$, respectively. We use $(n,p,q)=(2000,18,18)$, $(3000,26,26)$, and $(4000,36,36)$ so that dimension and sample size increase proportionally. For each setting, SW uses $N_X=2000$ and either $N_Y=0$ or $N_Y=2000$. Similar to \citet{KesslerLevina2023}, we evaluate performance at two coordinates of each leading canonical direction: the first coordinate, which is nonzero and used to assess power, and the last coordinate, which is zero and used to assess empirical size.

\begin{table}[!ht]
\centering
\setlength{\tabcolsep}{4pt}
\renewcommand{\cellalign}{lc}
\caption{Empirical size and power (\%) at significance level $\alpha = 0.05$ under rank-two model. Entries are Monte Carlo means over 500 replications. $\mathrm{SW}_X$ uses $(N_X,N_Y)=(2000,0)$ and $\mathrm{SW}_{XY}$ uses $(2000,2000)$.}
\label{tab:simulation-compact}
\begin{tabular}{@{}lllrrrr@{}}
\toprule
Setting & Side & Metric & PW & SW$_X$ & SW$_{XY}$ & CB\\
\midrule
\multirow{4}{*}{\makecell{$n=2000$\\ $p=q=18$ \\$s=9$}}
& X & Size & 5.6 & 5.2 & 5.2 & 5.4\\
&   & Power          & 33.2 & 34.2 & 36.6 & 32.4\\
& Y & Size & 5.2 & 5.2 & 5.2 & 4.8\\
&   & Power          & 31.2 & 33.8 & 36.0 & 30.2\\
\addlinespace[1pt]
 \hline
\addlinespace[1pt]
\multirow{4}{*}{\makecell{$n=3000$\\ $p=q=26$ \\$s=13$}}
& X & Size & 4.6 & 4.8 & 4.8 & 4.6\\
&   & Power          & 32.2 & 33.0 & 35.6 & 31.4\\
& Y & Size & 4.6 & 4.6 & 4.0 & 4.6\\
&   & Power          & 30.2 & 32.8 & 33.0 & 30.0\\
\addlinespace[1pt]
 \hline
\addlinespace[1pt]
\multirow{4}{*}{\makecell{$n=4000$\\ $p=q=36$ \\$s=18$}}
& X & Size & 5.0 & 5.6 & 4.4 & 5.2\\
&   & Power          & 28.8 & 28.6 & 29.2 & 28.2\\
& Y & Size & 4.4 & 4.4 & 4.2 & 4.8\\
&   & Power          & 29.0 & 29.4 & 31.2 & 28.4\\

\bottomrule
\end{tabular}
\end{table}

Table~\ref{tab:simulation-compact} reports empirical size and power across the simulation settings. All proposed methods and CB maintain empirical size close to the nominal $5\%$ level. PW and CB exhibit similar power, while SW achieves higher power than both methods with comparable empirical size for the $X$- and $Y$-side leading canonical directions. Interestingly, additional $X^U$ observations alone also improve the power of inference on the $Y$-side leading direction $v_1$, illustrating that unpaired observations from one view can improve inference on the other through the dependence between the two views. The power gains are most pronounced when $N_X=N_Y=2000$, that is, when additional observations from both views are available, demonstrating the benefit of incorporating unpaired data through SW.

\subsection{Increasing unpaired sample sizes}\label{sec:auxiliary-sizes}
In this section, we study the effect of increasing unpaired observations. We fix $(n,p,q)=(3000, 26,26)$, and set $N_X = m, N_Y \in \{m, 0\}$ for $m\in\{2000,3000,4000, 5000\}$. The first allocation increases only the $X^U$ sample, and the second increases both marginal samples. 

\begin{table}[!ht]\centering
\caption{Empirical size and power (\%) at significance level $\alpha = 0.05$ as unpaired sample sizes increases. Entries are Monte Carlo means over 500 replications.}\label{tab:simulation-growth}\label{tab:simulation-auxiliary}
\begin{tabular*}{\columnwidth}{@{\extracolsep{\fill}}rr|rrrr@{}}
\toprule
& & \multicolumn{2}{c}{$X$} & \multicolumn{2}{c}{$Y$}\\
\cmidrule(lr){3-4}\cmidrule(lr){5-6}
$N_X$ & $N_Y$ & Size & Power & Size & Power\\
\midrule
2000 & 0 & 4.8 & 33.0 & 4.6 & 32.8\\
2000 & 2000 & 4.8 & 35.6 & 4.0 & 33.0\\
\addlinespace[3pt]
3000 & 0 & 4.2 & 33.8 & 4.4 & 34.2\\
3000 & 3000 & 4.6 & 34.0 & 4.2 & 34.2\\
\addlinespace[3pt]
4000 & 0 & 3.8 & 34.2 & 4.4 & 33.8\\
4000 & 4000 & 4.2 & 35.6 & 4.0 & 34.6\\
\addlinespace[3pt]
5000 & 0 & 4.2 & 35.2 & 4.4 & 34.4\\
5000 & 5000 & 3.8 & 37.0 & 4.8 & 34.8\\
\bottomrule
\end{tabular*}
\end{table}

As shown in Table~\ref{tab:simulation-auxiliary}, increasing the number of marginal samples can further improve power while maintaining empirical size near the nominal $5\%$ level. When $N_Y=0$, increasing $N_X$ from 2000 to 5000 increases power from $33.0\%$ to $35.2\%$ for $X$ and from $32.8\%$ to $34.4\%$ for $Y$. When both marginal sample sizes are increased from 2000 to 5000, power rises from $35.6\%$ to $37.0\%$ for $X$ and from $33.0\%$ to $34.8\%$ for $Y$.

\subsection{Data-based simulation}

\begin{table}[!ht]\centering
\caption{Empirical size and power (\%) at significance level $\alpha = 0.05$ for data-based simulation. Entries are Monte Carlo means over 500 replications.}
\label{tab:simulation-five-cohort}
\begin{tabular*}{\columnwidth}{@{\extracolsep{\fill}}lllrrr@{}}
\toprule
Setting & Side & Metric & PW & SW & CB\\
\midrule
\multirow{4}{*}{Gaussian}
 & X & Size  & 3.8 & 3.8 & 2.6\\
 &   & Power & 14.6 & 14.8 & 11.8\\
 & Y & Size  & 5.4 & 5.8 & 2.0\\
 &   & Power & 88.8 & 89.0 & 80.6\\
\addlinespace[1pt]
 \hline
\addlinespace[1pt]
\multirow{4}{*}{Uniform}
 & X & Size  & 6.2 & 6.6 & 4.2\\
 &   & Power & 18.6 & 18.2 & 14.4\\
 & Y & Size  & 6.4 & 6.4 & 2.4\\
 &   & Power & 92.2 & 92.4 & 83.0\\
\addlinespace[1pt]
 \hline
\addlinespace[1pt]
\multirow{4}{*}{\shortstack[l]{Standardized\\Gamma(10)}}
 & X & Size  & 5.0 & 4.6 & 3.6\\
 &   & Power & 14.2 & 14.4 & 12.2\\
 & Y & Size  & 5.8 & 5.8 & 2.0\\
 &   & Power & 89.0 & 89.0 & 80.6\\
\addlinespace[1pt]
 \hline
\addlinespace[1pt]
\multirow{4}{*}{\shortstack[l]{Centered\\exponential}}
 & X & Size  & 4.8 & 4.6 & 3.6\\
 &   & Power & 15.8 & 15.2 & 12.6\\
 & Y & Size  & 6.8 & 6.8 & 2.6\\
 &   & Power & 90.6 & 90.8 & 82.6\\
 \hline
\addlinespace[2pt]
\multirow{4}{*}{Standardized $t_5$}
 & X & Size  & 6.0 & 5.8 & 4.2\\
 &   & Power & 14.4 & 14.2 & 10.4\\
 & Y & Size  & 6.6 & 6.8 & 2.6\\
 &   & Power & 88.0 & 88.0 & 77.6\\
\bottomrule
\end{tabular*}
\end{table}
For a more realistic setting, we directly use covariance matrices based on a neuroimaging dataset that we describe in Section~\ref{sec:real-data}. In this dataset, the dimensions are $p=21, q=4$ and sample sizes are $n=3056$, $N_X=214$, $N_Y=1762$. We use estimated covariance matrix $\hat{\Sigma}$ from real data which we then treat as ground truth to obtain canonical directions from CCA estimates. Since the estimated leading canonical directions are dense, we set the eleven coordinates with the smallest magnitudes in $\widehat u_1$ and the two coordinates with the smallest magnitudes in $\widehat v_1$ to zero. We normalize these modified vectors in their respective marginal covariance metrics and orthogonalize the remaining canonical directions against them and one another. We then reconstruct the cross-covariance matrix while preserving the estimated marginal covariance matrices and canonical correlations. The resulting covariance matrix $\Sigma$ therefore has the modified vectors as its leading canonical directions, satisfying \eqref{eq:cca-est}. We use the coordinate with smallest original magnitude among the zeroed coordinates to evaluate empirical size, and use the coordinate with smallest original magnitude among the retained non-zero coordinates to evaluate power. 

Since the real data may not be Gaussian, we also consider non-Gaussian settings. Let $L$ be the Cholesky factor of $\Sigma$. We generate $(X^\top,Y^\top)^\top=L\varepsilon$, where the coordinates of $\varepsilon$ are independent with mean zero and variance one. We use five latent variables: standard normal $N(0,1)$, $\operatorname{Unif}(-\sqrt{3},\sqrt{3})$, $(G-10)/\sqrt{10}$ with $G\sim\operatorname{Gamma}(10,1)$, $E-1$ with $E\sim\operatorname{Exp}(1)$, and $\sqrt{3/5}\,T$ with $T\sim t_5$. For each setting, paired and unpaired observations are generated independently from the same joint latent model, with only the corresponding view retained for each unpaired sample. This construction ensures that the paired and unpaired observations have matching marginal distributions.

Table~\ref{tab:simulation-five-cohort} shows that PW and SW perform similarly, with empirical sizes close to the nominal level, while SW provides modest power gains in some cases. One possible explanation is that the paired sample size is relatively large ($n=3056$) compared with the available unpaired sample sizes, leaving less room for the additional unpaired observations to improve efficiency. Consequently, SW provides only limited power improvement over PW in this experiment. Importantly, despite the departure from Gaussianity, the empirical sizes of the proposed methods remain close to the nominal level, suggesting that the procedures are reasonably robust in this setting. In comparison, CB tends to be conservative, with empirical sizes generally below the nominal level.

\subsection{Runtime comparison}\label{sec:runtime}
We compare the computational efficiency of our proposed methods with CB. All methods were run on a MacBook Pro with an Apple M1 Pro CPU and 16 GB of memory. Additionally, we also use $B = 500, 1000, 2000$ for CB with and without parallelization. As shown in Table~\ref{tab:direction-timing-workers}, our methods are extremely efficient since we use plug-in estimates. On the other hand, resampling methods like CB are computationally heavy. Even with eight parallel workers, the average runtime of CB at \(B=500\) is approximately 171 times that of SW using one worker.
\begin{table}[!ht]
\centering
\setlength{\tabcolsep}{4pt}
\caption{Mean $\pm$ SD of runtime (ms) over 500 replications at $n=3000$, $p=q=26$. SW uses $N_X=N_Y=2000$.}
\label{tab:direction-timing-workers}
\begin{tabular}{@{}lcr@{}}
\toprule
Method & Num.\ workers & Time (ms) \\
\midrule
PW & 1 & $0.743\pm0.078$ \\
SW & 1 & $0.879\pm0.050$ \\
\midrule
\multirow{3}{*}{CB ($B=500$)}
 & 1 & $437.795\pm25.217$ \\
 & 4 & $171.598\pm21.347$ \\
 & 8 & $150.185\pm15.137$ \\
\hline
\multirow{3}{*}{CB ($B=1000$)}
 & 1 & $864.012\pm39.309$ \\
 & 4 & $336.330\pm31.032$ \\
 & 8 & $291.912\pm20.582$ \\
\hline
\multirow{3}{*}{CB ($B=2000$)}
 & 1 & $1715.172\pm56.948$ \\
 & 4 & $670.601\pm52.737$ \\
 & 8 & $574.048\pm33.976$ \\
\bottomrule
\end{tabular}
\end{table}

\section{Brain-behavior association studies}\label{sec:real-data}
\label{sec:five-cohort-functional}
We apply our proposed methods to study the associations between resting-state functional connectivity and dimensions of psychopathology using the Reproducible Brain Charts (RBC) dataset, which is an open data resource provided by \citet{ShafieiEtAl2025}. The authors combine data from 5 large studies of brain development with participants aged between 6 and 22 years old. The dataset includes $n=3056$ imaging-behavior pairs, $N_X=214$ imaging-only observations, and $N_Y=1762$ behavioral-only observations. Only resting-state fMRI scans passing quality control are used to obtain imaging features. We refer the readers to the original paper of \citet{ShafieiEtAl2025} for more details in data processing and harmonization. The seven brain functional networks were defined by \citet{yeo2011organization}. We use the preprocessed imaging view $X\in\mathbb{R}^{21}$ that holds between-network functional connectivities and the behavioral view $Y\in\mathbb{R}^{4}$ that holds four harmonized behavioral scores: general psychopathology, internalizing, externalizing, and attention/hyperactivity \citep{HoffmannEtAl2023}.

\begin{table}[!h]\centering
\setlength{\tabcolsep}{1.5pt}
\caption{Brain-behavior leading canonical directions and corresponding $p$-values for PW, CB, and SW for the RBC data. A ${}^\star$ indicates $p<0.05$.}
\label{tab:five-cohort-coordinate-tests}
\begin{tabular}{@{}lrrr@{}}\toprule Coordinate & PW & CB & SW\\\midrule

\multicolumn{4}{@{}l}{\textit{Behavioral features}}\\

General psychopathology & $<0.001^{\star}$ & $0.003^{\star}$ & $<0.001^{\star}$\\

Internalizing & $0.215$ & $0.255$ & $0.257$\\

Externalizing & $<0.001^{\star}$ & $0.006^{\star}$ & $<0.001^{\star}$\\

Attention/hyperactivity & $<0.001^{\star}$ & $<0.001^{\star}$ & $<0.001^{\star}$\\

\addlinespace[2pt]
\hline

\multicolumn{4}{@{}l}{\textit{Connectivity features}}\\

Control--Default & $0.011^{\star}$ & $0.023^{\star}$ & $0.014^{\star}$\\

Control--DorsAttn & $0.612$ & $0.586$ & $0.668$\\

Control--Limbic & $0.866$ & $0.872$ & $0.904$\\

Control--SalVentAttn & $0.146$ & $0.167$ & $0.153$\\

Control--SomMot & $0.091$ & $0.114$ & $0.097$\\

Control--Visual & $0.233$ & $0.260$ & $0.220$\\

Default--DorsAttn & $0.691$ & $0.693$ & $0.661$\\

Default--Limbic & $0.060$ & $0.079$ & $0.070$\\

Default--SalVentAttn & $0.069$ & $0.094$ & $0.049^{\star}$\\

Default--SomMot & $0.328$ & $0.348$ & $0.376$\\

Default--Visual & $0.807$ & $0.819$ & $0.890$\\

DorsAttn--Limbic & $0.906$ & $0.950$ & $0.915$\\

DorsAttn--SalVentAttn & $0.858$ & $0.866$ & $0.977$\\

DorsAttn--SomMot & $0.010^{\star}$ & $0.015^{\star}$ & $0.011^{\star}$\\

DorsAttn--Visual & $0.586$ & $0.592$ & $0.614$\\

Limbic--SalVentAttn & $0.437$ & $0.454$ & $0.489$\\

Limbic--SomMot & $0.691$ & $0.685$ & $0.617$\\

Limbic--Visual & $0.323$ & $0.384$ & $0.345$\\

SalVentAttn--SomMot & $0.756$ & $0.746$ & $0.727$\\

SalVentAttn--Visual & $0.163$ & $0.178$ & $0.175$\\

SomMot--Visual & $0.722$ & $0.758$ & $0.678$\\

\bottomrule\end{tabular}
\par\smallskip{\scriptsize DorsAttn: Dorsal Attention; SalVentAttn: Salience/Ventral Attention; SomMot: Somatomotor.}
\end{table}

Table~\ref{tab:five-cohort-coordinate-tests} shows $p$-values obtained from PW, SW and CB. At the significant level $\alpha = 0.05$, all methods reject three null hypotheses for behavioral direction (general psychopathology, externalizing, and attention/hyperactivity) and two null hypotheses for functional connectivity direction (Control-Default and Dorsal Attention-Somatomotor). Previous works in neuroscience have reported significant associations between general psychopathology (the P factor) and resting-state functional connectivity between the default mode (DMN) and control networks \citep{xia2018linked,sripada2021widespread}. Moreover, some studies have linked attention deficits and disruptive behaviors to functional connectivity involving attention networks \citep{wong2021common}, and reported connectivity differences involving the somatomotor and dorsal attention networks in children with ADHD \citep{saad2022intrinsic}. These findings are broadly consistent with the contributions of attention/hyperactivity and externalizing scores in the leading behavioral canonical direction identified by our analysis. Interestingly, SW additionally determines the connectivity between DMN and salience/ventral attention networks as significant. This functional connectivity was shown to be associated with ADHD diagnosis by \citet{norman2023evidence}. Overall, these findings illustrate the usefulness of our proposed methods to brain-behavior association studies.

\section{Discussion}\label{sec:discussion}
In this work, we propose a one-step estimator and a sign-invariant Wald test for coefficients of the leading canonical directions with partially paired data. We show that the proposed test is asymptotically valid when $d=o(n)$, for arbitrary unpaired sample sizes. In addition, we extend our approach to construct confidence intervals for the leading canonical correlation. Simulations and a neuroimaging application demonstrate the favorable performance of our proposed methods.

Several extensions are of interest. One is to construct confidence intervals for canonical direction coefficients when the dimension grows with sample size. It would be interesting to see whether a jackknife correction similar to that in \citet{lin2024worthwhile} can be applied to canonical direction coefficients. Another future direction is to allow the unpaired samples to have different marginal distributions from the paired sample through specified density-ratio models \citep{LiGilbertDuanLuedtke2025}. This would require accounting for density-ratio estimation in both the augmentation and the variance estimator. We leave these questions for future work.

\section*{AI Use Statement}
In this work, we used generative AI tools for improving mathematical proofs, implementing numerical experiments, and improving the exposition and organization of the manuscript. We have reviewed, edited and verified all AI-assisted work. We take responsibility for the final content of the work, including all text, mathematical claims, empirical results, and artifacts prepared with this assistance.

\bibliographystyle{plainnat}
\bibliography{references}

\appendix
\section{Technical lemmas and derivations}
\label{sec:technical-tools}\label{sec:probabilistic-preliminaries}\label{sec:proof}
Define
\[
 \Psi_i=Z_iZ_i^\top-\Sigma,\qquad
 \Delta=\widehat\Sigma-\Sigma=\frac1n\sum_{i=1}^n\Psi_i,\qquad
 U_W=\frac1{n+N_W}\sum_{r=1}^{N_W}
       \{W_r^U(W_r^U)^\top-\Sigma_W\},
\]
for $W=X,Y$. Subtracting the paired average from the pooled average gives the identity
\begin{equation}\label{eq:pool-difference-decomposition}
 D_W=U_W-\gamma_W\Delta_W.
\end{equation}
Assumption~\ref{ass:model} implies, for a fixed $c_0>0$,
\[
 c_0\le\lambda_1\le1-c_0,\qquad
 \lambda_1^2-\lambda_2^2\ge c_0.
\]
The same bounds hold on a fixed operator-norm neighborhood of $\Sigma$. For theoretical analysis, we assume $(u_1^*)^\top\Sigma_Xu_1(S)>0$ and $(v_1^*)^\top\Sigma_Y v_1(S)>0$. The implemented statistics are independent of this orientation.

\subsection{Probabilistic tools}
\label{sec:probabilistic-tools}

\subsubsection{Conditional expectation and Gaussian moments}
\label{sec:prelim-gaussian-moments}

For $V\in L^2$ and a sigma-field $\mathcal F$, we use the contraction
and projection properties of conditional expectation
\citep[Theorems~4.1.11 and~4.1.15]{Durrett2019}
\begin{equation}\label{eq:prelim-conditional-expectation}
 \E[\E(V\mid\mathcal F)^2]\le\E V^2,\qquad
 \E[\{V-\E(V\mid\mathcal F)\}H]=0\quad(H\in L^2(\mathcal F)).
\end{equation}
Conditional Markov's inequality gives the form of Chebyshev used below
\begin{equation}\label{eq:prelim-conditional-chebyshev}
 \Pr\{|V-\E(V\mid\mathcal F)|>t\mid\mathcal F\}
 \le t^{-2}\Var(V\mid\mathcal F),\qquad t>0.
\end{equation}

\begin{lemma}[Gaussian regression, \citealp{Anderson2003}]
\label{lem:prelim-gaussian-regression}
Let $(X^\top,Y^\top)^\top$ be centered Gaussian with positive-definite covariance $S=\left(\begin{smallmatrix}A&C\\C^\top&B\end{smallmatrix}\right)$. Then
\begin{equation}\label{eq:prelim-gaussian-regression}
\begin{aligned}
 \E(Y\mid X)&=C^\top A^{-1}X,&
 \Cov(Y\mid X)&=B-C^\top A^{-1}C,\\
 \E(X\mid Y)&=CB^{-1}Y,&
 \Cov(X\mid Y)&=A-CB^{-1}C^\top.
\end{aligned}
\end{equation}
The residual $Y-C^\top A^{-1}X$ is independent of $X$, and conversely with the two views exchanged.
\end{lemma}

\begin{lemma}[Gaussian product moments, \citealp{Isserlis1918}]
\label{lem:prelim-gaussian-moments}
For centered jointly Gaussian scalars $V_1,\ldots,V_{2k}$,
\begin{equation}\label{eq:prelim-isserlis}
 \E\prod_{i=1}^{2k}V_i=\sum_{\pi}\prod_{\{a,b\}\in\pi}\E(V_aV_b),
\end{equation}
where $\pi$ ranges over pair partitions. Products of an odd number of centered jointly Gaussian variables have mean zero. For $G\sim N(0,\Gamma)$ and deterministic symmetric $M,N$, let $Q_M=G^\top MG-\tr(M\Gamma)$. Then
\begin{equation}\label{eq:prelim-quadratic-covariance}
 \E Q_M=0,\qquad \Cov(Q_M,Q_N)=2\tr(M\Gamma N\Gamma),\qquad
 \Var(Q_M)=2\|\Gamma^{1/2}M\Gamma^{1/2}\|_{\Frob}^2.
\end{equation}
If $\theta_r$ are the eigenvalues of $\Gamma^{1/2}M\Gamma^{1/2}$, then
\begin{equation}\label{eq:gaussian-fourth-moment}
 \E Q_M^4=12\Big(\sum_r\theta_r^2\Big)^2+48\sum_r\theta_r^4
                  \le15\{\Var(Q_M)\}^2.
\end{equation}
\end{lemma}
The quadratic-form covariance identity is the fourth-order case of
\eqref{eq:prelim-isserlis}. For the last display, diagonalization gives
$Q_M\overset d=\sum_r\theta_r(\zeta_r^2-1)$ with independent standard
normals. Since $\E(\zeta_r^2-1)^2=2$ and $\E(\zeta_r^2-1)^4=60$,
the surviving terms are $60\sum_r\theta_r^4+
24\sum_{r<s}\theta_r^2\theta_s^2$, which gives the stated formula.

\begin{lemma}[Gaussian covariance increments]
\label{lem:gaussian-covariance-increments}
Let $Z\sim N(0,\Sigma)$ and $\Psi=ZZ^\top-\Sigma$. For deterministic
vectors $a,b,c,e$ of matching dimension,
\begin{equation}\label{eq:gaussian-covariance-increments}
 \E\{(a^\top\Psi b)(c^\top\Psi e)\}
 =(a^\top\Sigma c)(b^\top\Sigma e)
 +(a^\top\Sigma e)(b^\top\Sigma c).
\end{equation}
In particular,
\begin{equation}\label{eq:gaussian-covariance-increment-variance}
 \E(a^\top\Psi b)^2
 =(a^\top\Sigma a)(b^\top\Sigma b)+(a^\top\Sigma b)^2.
\end{equation}
\end{lemma}

\begin{proof}
Expanding both centered bilinear forms gives
\[
 \E\{(a^\top Z)(b^\top Z)(c^\top Z)(e^\top Z)\}
 -(a^\top\Sigma b)(c^\top\Sigma e).
\]
Of the three pairings in \eqref{eq:prelim-isserlis}, one equals
$(a^\top\Sigma b)(c^\top\Sigma e)$ and cancels the centering term.
The remaining two give \eqref{eq:gaussian-covariance-increments}.
Setting $(c,e)=(a,b)$ gives
\eqref{eq:gaussian-covariance-increment-variance}.
\end{proof}

\subsubsection{Gaussian sample covariance}
\label{sec:prelim-covariance}

\begin{lemma}[Covariance tails and moments]
\label{lem:prelim-covariance}
For independent $Z_i\sim N(0,\Sigma)$ in $\R^d$ with $\Sigma\succ0$,
\begin{equation}\label{eq:cov-conc}
\begin{aligned}
 &\Pr\left\{\|\Delta\|_{\op}>
 C\|\Sigma\|_{\op}
 \left(\sqrt{\frac{d+t}{n}}+\frac{d+t}{n}\right)\right\}\le e^{-t},\\
 &(\E\|\Delta\|_{\op}^k)^{1/k}
 \le C_k\|\Sigma\|_{\op}\{\sqrt{d/n}+d/n\},
\end{aligned}
\end{equation}
for $t\ge1$ and each fixed $k\ge1$.
\end{lemma}
These standard sample covariance bounds follow from \citet[Theorem~6.5]{wainwright2019high}. In particular, under Assumption~\ref{ass:model} and $d=o(n)$,
\[
(\E\|\Delta\|_{\op}^k)^{1/k}
=O(\sqrt{d/n})
\]
for every fixed $k\ge1$.

\subsubsection{Brascamp--Lieb inequality and a conditional Wishart bound}
\label{sec:prelim-brascamp-lieb}

\begin{lemma}[Brascamp--Lieb inequality on a smooth convex domain]
\label{lem:brascamp-lieb}
Let $\Omega$ be a bounded convex domain in a finite-dimensional Euclidean
space with $C^2$ boundary. Suppose $V$ is $C^2$ on a neighborhood of
$\overline\Omega$ and $\nabla^2V$ is positive definite there. For the
probability measure $\mu(dx)\propto e^{-V(x)}\mathbf1_\Omega(x)dx$
and any $h$ that is $C^1$ on a neighborhood of $\overline\Omega$,
\begin{equation}\label{eq:prelim-brascamp-lieb}
 \Var_\mu(h)\le
 \E_\mu\langle\nabla h,(\nabla^2V)^{-1}\nabla h\rangle.
\end{equation}
In particular, if $\nabla^2V\succeq\rho I$, then
$\Var_\mu(h)\le\rho^{-1}\E_\mu\|\nabla h\|^2$.
\end{lemma}
This is the Euclidean, $N=\infty$ case of
\citet[Theorem~1.2(1)]{KolesnikovMilman2017}, extending the variance
inequality of \citet{BrascampLieb1976} to convex domains. No boundary
condition on $h$ is imposed. We use the Frobenius Euclidean structure
$\langle H,K\rangle=\tr(HK)$ on the symmetric matrices.

\begin{lemma}[Conditional Wishart variance]
\label{lem:conditional-wishart}
Let $n>d+1$, $\Sigma\succ0$, and $0<\delta<\lambda_{\min}(\Sigma)$.
Define
\begin{equation}\label{eq:local-good-event}
 \Omega_\delta=\{S=S^\top:\|S-\Sigma\|_{\op}<\delta\},\qquad
 \mathcal E_n=\{\widehat\Sigma\in\Omega_\delta\}.
\end{equation}
For any real $h$ that is $C^1$ on a neighborhood of
$\overline{\Omega_\delta}$,
\begin{equation}\label{eq:conditional-wishart-variance}
\begin{aligned}
 \Var\{h(\widehat\Sigma)\mid\mathcal E_n\}
 &\le\frac{2}{n-d-1}
   \E\{\|\widehat\Sigma^{1/2}\nabla h(\widehat\Sigma)
                 \widehat\Sigma^{1/2}\|_{\Frob}^2\mid\mathcal E_n\}\le\frac{2(\|\Sigma\|_{\op}+\delta)^2}{n-d-1}
              \E\{\|\nabla h(\widehat\Sigma)\|_{\Frob}^2\mid\mathcal E_n\}.
\end{aligned}
\end{equation}
\end{lemma}

\begin{proof}
The Wishart density \citep{Anderson2003} of $S=\widehat\Sigma$ is
proportional, on the positive-definite cone, to
$\exp\{-V(S)\}$, where
\[
 V(S)=\frac n2\tr(\Sigma^{-1}S)-\frac{n-d-1}{2}\log\det S.
\]
The degrees of freedom are $n$ because the known zero mean is not estimated. Differentiation gives
\[
\begin{aligned}
 DV(S)[H]&=\tfrac n2\tr(\Sigma^{-1}H)-\tfrac{n-d-1}{2}\tr(S^{-1}H),\\
 D^2V(S)[H,K]&=\tfrac{n-d-1}{2}\tr(S^{-1}HS^{-1}K).
\end{aligned}
\]
Thus the Hessian operator maps $H$ to $(n-d-1)S^{-1}HS^{-1}/2$. Its inverse maps $H$ to
$2SHS/(n-d-1)$. It is positive definite since $n>d+1$. In particular,
\[
 D^2V(S)[H,H]=\tfrac{n-d-1}{2}\|S^{-1/2}HS^{-1/2}\|_{\Frob}^2
 \ge\frac{n-d-1}{2(\|\Sigma\|_{\op}+\delta)^2}\|H\|_{\Frob}^2
 \quad(S\in\overline{\Omega_\delta}).
\]
To handle the nonsmooth operator-norm boundary, for integers $k\ge1$ use
\[
 \Omega_{\delta,k}=
 \{S=S^\top:\tr((S-\Sigma)^{2k})<\delta^{2k}\}.
\]
These are convex Schatten $2k$-norm balls. Their boundaries are smooth: the defining polynomial has gradient $2k(S-\Sigma)^{2k-1}\ne0$ on the boundary. Their closures are contained in the positive-definite cone, and $\Omega_{\delta,k}\uparrow\Omega_\delta$, since Schatten $2k$ norms decrease to the operator norm. Apply Lemma~\ref{lem:brascamp-lieb} on each $\Omega_{\delta,k}$ with its normalized restricted Wishart density. The inverse Hessian gives
\[
 \langle\nabla h,(\nabla^2V)^{-1}\nabla h\rangle
 =\frac{2}{n-d-1}\tr\{\nabla h\,S\,\nabla h\,S\}
 =\frac{2}{n-d-1}\|S^{1/2}\nabla h\,S^{1/2}\|_{\Frob}^2.
\]
Both $h$ and its gradient are bounded on $\overline{\Omega_\delta}$ for each fixed $n,d$. Dominated convergence of the conditional integrals as $k\to\infty$ proves the first inequality. The constant is independent of $k$. The bound $\|S^{1/2}GS^{1/2}\|_{\Frob}\le\|S\|_{\op}\|G\|_{\Frob}$ gives the second. The Euclidean space has dimension $d(d+1)/2$, but no factor of that dimension enters the inequality.
\end{proof}

\subsubsection{Independent marginal samples}
\label{sec:prelim-marginal-samples}
Let $G_W$ be a symmetric matrix measurable with respect to the paired sample and of the same dimension as $U_W$. Conditional on the paired sample, Lemma~\ref{lem:prelim-gaussian-moments} gives
\begin{equation}\label{eq:mixed-conditional-variance}
\begin{aligned}
 \E\{\langle G_W,U_W\rangle_{\Frob}\mid Z_1,\ldots,Z_n\}&=0,\\
 \Var\{\langle G_W,U_W\rangle_{\Frob}\mid Z_1,\ldots,Z_n\}
 &=\frac{2N_W}{(n+N_W)^2}
      \tr(G_W\Sigma_WG_W\Sigma_W) \le \frac Cn\|G_W\|_{\Frob}^2.
\end{aligned}
\end{equation}
Indeed, one summand is the centered Gaussian quadratic form $W^\top G_WW-\tr(G_W\Sigma_W)$, whose variance is $2\tr(G_W\Sigma_WG_W\Sigma_W)$. The last inequality uses $N_W/(n+N_W)^2\le1/(4n)$ and bounded $\|\Sigma_W\|_{\op}$. The formula also covers $N_W=0$.
Consequently,
\begin{equation}\label{eq:mixed-average-bound}
 \langle G_W,U_W\rangle_{\Frob}=O_p(r_n/\sqrt n)
 \quad\text{whenever}\quad \|G_W\|_{\Frob}=O_p(r_n).
\end{equation}
The same bound holds on a high-probability event determined by the paired sample. In particular, if
$\mathbf1_{\mathcal E_n}\|G_W\|_{\Frob}\le C\|\Delta\|_{\op}$, then under $d=o(n)$,
\begin{equation}\label{eq:localized-mixed-second-moment}
 \E\!\left[\{\mathbf1_{\mathcal E_n}
        \langle G_W,U_W\rangle_{\Frob}\}^2\right]
 \le\frac Cn\E\|\Delta\|_{\op}^2\le Cd/n^2.
\end{equation}
For any random $R_n$ and deterministic $r_n>0$,
\begin{equation}\label{eq:prelim-remove-event}
 \Pr(|R_n|>Mr_n)\le\Pr(\mathcal E_n^c)
                   +\Pr(\mathbf1_{\mathcal E_n}|R_n|>Mr_n).
\end{equation}
Since $\Pr(\mathcal E_n^c)\to0$, an $O_p(r_n)$ bound established on $\mathcal E_n$ also holds unconditionally. No bound on $G_W$ is needed outside $\mathcal E_n$.

\subsubsection{Lyapunov CLT and studentization}
\label{sec:prelim-clt}

\begin{lemma}[Lyapunov's theorem]\label{lem:prelim-lyapunov}
Let $V_{n,1},\ldots,V_{n,k_n}$ be independent centered random variables within for each $n$ and $s_n^2=\sum_r\E V_{n,r}^2\in(0,\infty)$.
If, for some $\eta>0$,
\begin{equation}\label{eq:prelim-lyapunov-condition}
 s_n^{-2-\eta}\sum_r\E|V_{n,r}|^{2+\eta}\longrightarrow0,
\end{equation}
then $s_n^{-1}\sum_rV_{n,r}\rightsquigarrow N(0,1)$.
\end{lemma}
This is the Lyapunov central limit theorem for triangular arrays \citep[Theorem~3.4.10 and Exercise~3.4.12]{Durrett2019}. For the three-sample averages in this paper, the summands are centered Gaussian quadratic forms of uniformly bounded variance, with weights $n^{-1},(n+N_X)^{-1},(n+N_Y)^{-1}$. By \eqref{eq:gaussian-fourth-moment}, the sum of fourth moments of the summands in $\sqrt{n}\mathcal L_n$ is bounded by
\begin{equation}\label{eq:common-lyapunov-bound}
 C\left\{\frac1n+\frac{N_Xn^2}{(n+N_X)^4}
                    +\frac{N_Yn^2}{(n+N_Y)^4}\right\}=O(n^{-1}).
\end{equation}
The paired contribution is $Cn^{-1}$. The $W$ contribution is $Cn^{-1}t/(1+t)^4\le C/n$, where $t=N_W/n$.
If the total variance is bounded below, this verifies Lyapunov's condition with $\eta=2$.

We also use Slutsky's theorem and continuous mapping theorem \citep[Theorem~3.2.10 and Exercises~3.2.12--3.2.14]{Durrett2019}:
if $A_n\rightsquigarrow N(0,1)$, $B_n\xrightarrow{p}1$ and $C_n\xrightarrow{p}0$, then
\begin{equation}\label{eq:prelim-studentization}
 (A_n+C_n)/B_n\rightsquigarrow N(0,1),\qquad \{(A_n+C_n)/B_n\}^2\rightsquigarrow\chi_1^2.
\end{equation}
No independence among $A_n,B_n,C_n$ is needed. The quotient can be assigned any fixed value on $B_n=0$, whose probability tends to zero.

\subsection{Optimal augmentation equations}\label{sec:generic-projection}

\subsubsection{Influence-function expansion and one-view augmentation}
\label{sec:augmentation-derivation}
We give the intermediate steps underlying Section~\ref{sec:opt-aug}. For a covariance matrix $S$ in a neighborhood of $\Sigma$, define
\[
 T_j(S)=e_j^\top u_1(S),
\]
using the sign convention stated above. Thus $T_j(\Sigma)=u_{1j}^*$ and $T_j(\widehat\Sigma)=\widetilde u_{1j}$, where the tilde indicates alignment with the population direction. The first-order Taylor expansion is
\[
 \widetilde u_{1j}-u_{1j}^*
 =\frac1n\sum_{i=1}^n g_j(X_i,Y_i)+R_{j,n},
 \qquad
 g_j(X,Y)=DT_j(\Sigma)[ZZ^\top-\Sigma],
\]
where $R_{j,n}$ is the remainder term. We first treat the augmentation functions as fixed. Lemma~\ref{lem:augmented-expansion} accounts for their estimation using the paired sample. Suppose $N_X>0$ and $N_Y=0$. For a fixed centered square-integrable function $a_j(X)$, write
\[
 \bar a_{j,\mathrm{paired}}=\frac1n\sum_{i=1}^n a_j(X_i),
 \qquad
 \bar a_{j,U}=\frac1{N_X}\sum_{r=1}^{N_X}a_j(X_r^U).
\]
Its pooled average satisfies $\bar a_{j,\mathrm{pooled}}=(1-\gamma_X)\bar a_{j,\mathrm{paired}} +\gamma_X\bar a_{j,U}$. Replacing the paired average by this pooled average gives
\[
 \widetilde u_{1j}^{\mathrm{Semi},X}(a_j)
 =\widetilde u_{1j}+\bar a_{j,\mathrm{pooled}}-\bar a_{j,\mathrm{paired}}
 =\widetilde u_{1j}+\gamma_X(\bar a_{j,U}-\bar a_{j,\mathrm{paired}}).
\]
Consequently,
\begin{equation}\label{eq:inf-decomp}
 \widetilde u_{1j}^{\mathrm{Semi},X}(a_j)-u_{1j}^*
 =\frac1n\sum_{i=1}^n\{g_j(X_i,Y_i)-\gamma_Xa_j(X_i)\}
   +\gamma_X\bar a_{j,U}+R_{j,n}.
\end{equation}
The variance of $\sqrt{n}\mathcal L_n$ is
\[
 V_{j,X}(a_j)
 =\Var\{g_j(X,Y)-\gamma_Xa_j(X)\}
  +\gamma_X(1-\gamma_X)\Var\{a_j(X)\}.
\]
It is minimized by $a_j^*(X)=\E\{g_j(X,Y)\mid X\}$, as follows from \eqref{eq:prelim-conditional-expectation} or the optimality equations we will show below.
\subsubsection{Two-view augmentation}
For $N_X,N_Y>0$ and fixed centered square-integrable functions $a_j(X),b_j(Y)$, define $\bar b_{j,U}$ and the paired and pooled $b_j$-averages analogously. Adding both pooled-minus-paired adjustments to $\widetilde u_{1j}$ gives
\begin{equation}\label{eq:inf-decomp-coupled}
\begin{aligned}
 \widetilde u_{1j}^{\mathrm{Semi}}(a_j,b_j)-u_{1j}^*
 ={}&\frac1n\sum_{i=1}^n
 \{g_j(X_i,Y_i)-\gamma_Xa_j(X_i)-\gamma_Yb_j(Y_i)\}\\
 &+\gamma_X\bar a_{j,U}+\gamma_Y\bar b_{j,U}+R_{j,n}.
\end{aligned}
\end{equation}
Independence of the three samples gives the variance \eqref{eq:main-augmentation-variance}. We use the spaces $L_0^2(P_{XY})$, $L_0^2(P_X)$, and $L_0^2(P_Y)$ defined in Proposition~\ref{prop:main-projection} to derive the optimality equations.

\begin{proof}[Proof of Proposition~\ref{prop:main-projection}]
For centered square-integrable $g_j,a_j,b_j$, expansion gives
\begin{equation}\label{eq:generic-variance}
\begin{aligned}
 V_j(a_j,b_j)={}&\Var(g_j)-2\gamma_X\Cov(g_j,a_j)
 -2\gamma_Y\Cov(g_j,b_j)\\
 &+\gamma_X\Var(a_j)+\gamma_Y\Var(b_j)
 +2\gamma_X\gamma_Y\Cov(a_j,b_j).
\end{aligned}
\end{equation}
The first variation vanishes for all centered perturbations precisely when, on the sides with positive sampling fraction,
\begin{equation}\label{eq:projection-equations}
 a_j+\gamma_Y\E(b_j\mid X)=\E(g_j\mid X),\qquad
 b_j+\gamma_X\E(a_j\mid Y)=\E(g_j\mid Y).
\end{equation}
Indeed, the $a_j$-variation is $2\gamma_X\E[\{a_j+\gamma_Y\E(b_j\mid X)-\E(g_j\mid X)\}\delta a_j]$, and the $b_j$-variation is analogous. On the centered $L^2$ spaces, define $P_{Y\to X}b=\E(b\mid X)$ and $P_{X\to Y}a=\E(a\mid Y)$. Both operators have norm at most one by \eqref{eq:prelim-conditional-expectation}. For $\gamma_X\gamma_Y<1$, elimination
in \eqref{eq:projection-equations} gives
\[
 a_j=(I-\gamma_X\gamma_YP_{Y\to X}P_{X\to Y})^{-1}
       \{\E(g_j\mid X)-\gamma_YP_{Y\to X}\E(g_j\mid Y)\},\qquad
 b_j=\E(g_j\mid Y)-\gamma_XP_{X\to Y}a_j.
\]
The inverse exists by its norm-convergent Neumann series. At this solution the exact excess variance for perturbations $h(X),k(Y)$ is
\begin{equation}\label{eq:excess-variance}
\begin{aligned}
 &\gamma_X\E h^2+\gamma_Y\E k^2+2\gamma_X\gamma_Y\E(hk) \ge(1-\sqrt{\gamma_X\gamma_Y})
                   (\gamma_X\E h^2+\gamma_Y\E k^2).
\end{aligned}
\end{equation}
Thus the solution is the unique minimizer when both fractions are positive. If $\gamma_X>0$ and $\gamma_Y=0$, the minimizing function is $a_j(X)=\E\{g_j(X,Y)\mid X\}$, and $b_j$ does not affect the variance. The corresponding statement holds with $X$ and $Y$ exchanged. Multiplying the normal equations by $a_j,b_j$ and taking expectations also gives the identity used in the variance calculations
\begin{equation}\label{eq:gain-identity}
\begin{aligned}
 \Var(g_j)-V_j(a_j,b_j)
 &=\gamma_X\Cov(g_j,a_j)+\gamma_Y\Cov(g_j,b_j)\\
 &=\gamma_X\Var(a_j)+\gamma_Y\Var(b_j)
                  +2\gamma_X\gamma_Y\Cov(a_j,b_j).
\end{aligned}
\end{equation}
\end{proof}

\begin{lemma}[Two-component augmentation and variance]
\label{lem:two-component-augmentation}
Let $g\in L_0^2(P_{XY})$, $\phi\in L_0^2(P_X)$, and $\psi\in L_0^2(P_Y)$, with $\E\phi^2=\E\psi^2=1$. Suppose
\begin{equation}\label{eq:two-component-conditionals}
\begin{aligned}
 \E(\psi\mid X)&=\rho\phi,& \E(\phi\mid Y)&=\rho\psi,\\
 \E(g\mid X)&=b_X\phi,& \E(g\mid Y)&=b_Y\psi,
\end{aligned}
\end{equation}
where $|\rho|<1$. Hold the law, $g,\phi,\psi,b_X,b_Y$, and $\rho$ fixed as $\gamma=(\gamma_X,\gamma_Y)$ varies in $[0,1]^2$. The functions $a^\gamma=\alpha_\gamma\phi$ and $b^\gamma=\beta_\gamma\psi$, with
\begin{equation}\label{eq:two-component-coefficients}
 \alpha_\gamma=\frac{b_X-\gamma_Y\rho b_Y}{1-\gamma_X\gamma_Y\rho^2},
 \qquad
 \beta_\gamma=\frac{b_Y-\gamma_X\rho b_X}{1-\gamma_X\gamma_Y\rho^2},
\end{equation}
solve the two projection equations and minimize the variance criterion \eqref{eq:main-augmentation-variance}, with $g$ in place of $g_j$. The minimum, denoted by $V_\gamma$, satisfies
\begin{equation}\label{eq:two-component-variance}
 \Var(g)-V_\gamma
 =\frac{\gamma_Xb_X^2+\gamma_Yb_Y^2
              -2\gamma_X\gamma_Y\rho b_Xb_Y}
             {1-\gamma_X\gamma_Y\rho^2},
\end{equation}
and
\begin{equation}\label{eq:two-component-monotonicity}
 \partial_{\gamma_X}V_\gamma=-\alpha_\gamma^2,
 \qquad \partial_{\gamma_Y}V_\gamma=-\beta_\gamma^2.
\end{equation}
\end{lemma}

\begin{proof}
The conditional identities give $\E(\phi\psi)=\rho$ and reduce the projection equations to
\[
 \begin{pmatrix}1&\gamma_Y\rho\\\gamma_X\rho&1\end{pmatrix}
 \begin{pmatrix}\alpha_\gamma\\\beta_\gamma\end{pmatrix}
 =\begin{pmatrix}b_X\\b_Y\end{pmatrix}.
\]
Its determinant is $1-\gamma_X\gamma_Y\rho^2>0$, and its inverse is
\[
 \frac1{1-\gamma_X\gamma_Y\rho^2}
 \begin{pmatrix}1&-\gamma_Y\rho\\-\gamma_X\rho&1\end{pmatrix}.
\]
This proves \eqref{eq:two-component-coefficients}. The excess-variance identity \eqref{eq:excess-variance} proves minimality. When $\gamma_X\gamma_Y<1$, each function whose sampling fraction is positive is uniquely determined at the minimum. At $(1,1)$ the displayed pair remains a minimizer, without a uniqueness claim over arbitrary marginal functions. Since
$\Cov(g,\phi)=b_X$ and $\Cov(g,\psi)=b_Y$, \eqref{eq:gain-identity} gives
\[
 \Var(g)-V_\gamma
 =\gamma_X b_X\alpha_\gamma+\gamma_Y b_Y\beta_\gamma,
\]
which yields \eqref{eq:two-component-variance} by substitution. To verify the derivative without differentiating the fitted functions, rewrite
\[
 \Var(g)-V_\gamma
 =\gamma_Yb_Y^2+
       \frac{\gamma_X(b_X-\gamma_Y\rho b_Y)^2}
            {1-\gamma_X\gamma_Y\rho^2}.
\]
Direct differentiation gives
\[
\frac{\partial G_\gamma}{\partial\gamma_X} =\alpha_\gamma^2\ge0, \qquad \frac{\partial G_\gamma}{\partial\gamma_Y} =\beta_\gamma^2\ge0.
\]
Thus, the variance reduction is nondecreasing in both $\gamma_X$ and $\gamma_Y$. Since $1-\gamma_X\gamma_Y\rho^2>0$, these identities extend continuously to the boundary.
\end{proof}

\subsection{Derivatives and augmentation formulas}
\label{sec:coordinate-calculations}

\subsubsection{Derivatives}
Let $S=\left(\begin{smallmatrix}A&C\\C^\top&B\end{smallmatrix}\right)$ be a covariance matrix in a small neighborhood of $\Sigma$. In this subsection, canonical quantities are evaluated at $S$ unless stated otherwise, and we suppress this dependence in the notation. In particular, $(u,v,\lambda)$ denotes its  leading canonical triplet, with $\lambda=\lambda_1(S)$
\begin{equation}\label{eq:cca-kkt}
 Cv=\lambda Au,\qquad C^\top u=\lambda Bv,\qquad
 u^\top Au=v^\top Bv=1.
\end{equation}
For a symmetric perturbation $H$, write $h_X=Du(S)[H]$, $h_Y=Dv(S)[H]$, and $L=D\lambda(S)[H]$. Differentiation gives
\begin{equation}\label{eq:cca-linear-equations}
\begin{aligned}
 &\lambda Ah_X-Ch_Y+LAu=H_{XY}v-\lambda H_Xu,\\
 &\lambda Bh_Y-C^\top h_X+LBv=H_{XY}^\top u-\lambda H_Yv,\\
 &u^\top Ah_X=-\tfrac12u^\top H_Xu,\qquad
 v^\top Bh_Y=-\tfrac12v^\top H_Yv.
\end{aligned}
\end{equation}
Taking the inner products of the first two equations with $u$ and $v$ and adding them yields
\begin{equation}\label{eq:lambda-linear-expansion}
 D\lambda(S)[H]=u^\top H_{XY}v -\frac\lambda2(u^\top H_Xu+v^\top H_Yv).
\end{equation}

\subsubsection{Reduced-resolvent representation of the estimator}
Define
\begin{equation}\label{eq:one-sided-cca-operator}
 K=A^{-1/2}CB^{-1/2},\qquad
 M=KK^\top=A^{-1/2}CB^{-1}C^\top A^{-1/2}.
\end{equation}
Let
\[
 a=\lambda^2,\qquad w=A^{1/2}u,\qquad z=B^{1/2}v,
 \qquad P=ww^\top.
\]
Then $\|w\|=\|z\|=1$, $Kz=\lambda w$, $K^\top w=\lambda z$, and $Mw=aw$. Assumption~\ref{ass:model} keeping the leading canonical correlation away from one ensure that
\begin{equation}\label{eq:reduced-resolvent-defs}
 R=\{aI-M+P\}^{-1}(I-P),\qquad
 J=(I-\gamma_X\gamma_YaM)^{-1}
\end{equation}
are well defined and uniformly bounded on a fixed neighborhood of the population covariance $\Sigma$. Let $w_1=w,w_2,\ldots,w_p$ be an orthonormal eigenbasis of $M$, with eigenvalues $\lambda_\ell^2=\lambda_\ell^2(S)$ and $\lambda_\ell=0$ for $\ell>m$. Then
\begin{equation}\label{eq:reduced-resolvent-spectrum}
 Rw_1=0,\qquad
 Rw_\ell=(a-\lambda_\ell^2)^{-1}w_\ell\quad(\ell\ge2),
 \qquad
 Jw_\ell=(1-\gamma_X\gamma_Ya\lambda_\ell^2)^{-1}w_\ell.
\end{equation}
Thus $R$ collects all non-leading eigendirections without choosing or differentiating them individually.

For a fixed target $c\in\R^p$, put
\begin{equation}\label{eq:reduced-target}
 r_c=r_c(S)=A^{-1/2}c,
 \qquad \theta_c(S)=c^\top u=r_c^\top w.
\end{equation}
Set $\eta=\gamma_X\gamma_Y$ and define the symmetric spectral matrices
\begin{equation}\label{eq:reduced-augmentation-spectral}
\begin{aligned}
 G_X={}&-\frac{1-\gamma_Ya^2}{2(1-\eta a^2)}P
 +(1-a)(1-\gamma_Ya)MRJ,\\
 G_Y={}&-\frac{1-\gamma_X}{2(1-\eta a^2)}P
 +(1-a)(I-\gamma_XM)RJ.
\end{aligned}
\end{equation}
For a matrix $H$, write $\operatorname{sym}(H)=(H+H^\top)/2$, and define
\begin{equation}\label{eq:coordinate-augmentation-matrices}
\begin{aligned}
 B_{X,c}(S)
 &=\operatorname{sym}\!\left(A^{-1/2}wr_c^\top G_XA^{-1/2}\right),\\
 B_{Y,c}(S)
 &=\operatorname{sym}\!\left(B^{-1/2}K^\top wr_c^\top G_YKB^{-1/2}\right).
\end{aligned}
\end{equation}
Since $D_X,D_Y$ are symmetric,
\begin{equation}\label{eq:coordinate-matrix-corrections}
\begin{aligned}
 \langle B_{X,c}(S),D_X\rangle_{\Frob}
 &=r_c^\top G_XA^{-1/2}D_XA^{-1/2}w,\\
 \langle B_{Y,c}(S),D_Y\rangle_{\Frob}
 &=r_c^\top G_YKB^{-1/2}D_YB^{-1/2}K^\top w.
\end{aligned}
\end{equation}
Expanding these identities in the eigenbasis of $M$ gives the corrections in \eqref{eq:shared-coordinate-correction}. Indeed,
$r_c^\top w_\ell=c^\top u_\ell$, and for $\ell\ge2$ the $X$ coefficient is
\[
 \frac{(1-a)\lambda_\ell^2(1-\gamma_Ya)}
      {(a-\lambda_\ell^2)(1-\eta a\lambda_\ell^2)}.
\]
For $2\le\ell\le m$, let $z_\ell=B^{1/2}v_\ell$ be the corresponding right singular vector of $K$. Using
$K^\top w_\ell=\lambda_\ell z_\ell$, the $Y$ coefficient is
\[
 \frac{(1-a)\lambda\lambda_\ell(1-\gamma_X\lambda_\ell^2)}
      {(a-\lambda_\ell^2)(1-\eta a\lambda_\ell^2)}.
\]
The $P$ terms give the leading correction in \eqref{eq:shared-coordinate-correction}. If $\ell>m$, then $\lambda_\ell=0$ and both non-leading corrections vanish. Consequently, for any covariance matrix $S$ in the local neighborhood and any fixed $c\in\mathbb R^p$, the following matrix representation is equivalent to the estimator in $\eqref{eq:ss-estimator}$ when $c=e_j$
\begin{equation}\label{eq:ss-matrix-implementation}
 \widehat\theta_c^{\rm Semi}
 =\theta_c(\widehat\Sigma)
 +\langle B_{X,c}(\widehat\Sigma),D_X\rangle_{\Frob}
 +\langle B_{Y,c}(\widehat\Sigma),D_Y\rangle_{\Frob}.
\end{equation}
The representation does not depend on the choice of eigenvectors associated with repeated non-leading canonical correlations.

\subsubsection{Coordinate influence function and optimal augmentations}
We now introduce population canonical coordinates. Let
\[
 \Sigma_X^{-1/2}\Sigma_{XY}\Sigma_Y^{-1/2}=Q_X \Lambda Q_Y^\top,
 \qquad \Lambda=\diag_{p\times q}(\lambda_1,\ldots,\lambda_m),
\]
be a full singular value decomposition, where $Q_X\in\mathbb R^{p\times p}$ and $Q_Y\in\mathbb R^{q\times q}$ are orthogonal. Define
\begin{equation}\label{eq:fixed-canonical-transform}
 \mathcal W_X=Q_X^\top\Sigma_X^{-1/2},\qquad
 \mathcal W_Y=Q_Y^\top\Sigma_Y^{-1/2},\qquad
 \mathcal W=\diag(\mathcal W_X,\mathcal W_Y).
\end{equation}
The transformed population covariance $\mathcal W\Sigma\mathcal W^\top$ has blocks $I_p, I_q, \Lambda$. Within calculations in these fixed coordinates, we use the same symbols for the transformed observations, covariance matrices, and perturbations. The transformation itself does not depend on the sample. Here $\lambda_\ell$ denotes a population canonical correlation, and the leading directions are $e=e_1\in\mathbb R^p$ and $f=e_1\in\mathbb R^q$. An original coordinate target becomes the fixed vector
\begin{equation}\label{eq:transformed-test-functional}
 c_j=\mathcal W_Xe_j,
 \qquad c_{j\ell}:=(c_j)_\ell=e_j^\top u_\ell^*,
 \qquad \|c_j\|\le C.
\end{equation}
The scalar target is unchanged because $e_j^\top u=(\mathcal W_Xe_j)^\top(\mathcal W_X^{-\top}u)$. At the population covariance $\Sigma$, the derivative of $M$ in the direction $H$ is
\begin{equation}\label{eq:one-sided-M-derivative}
\begin{aligned}
DM(\Sigma)[H] = -\tfrac12 H_X\Lambda\Lambda^\top
-\tfrac12\Lambda\Lambda^\top H_X +H_{XY}\Lambda^\top
+\Lambda H_{XY}^\top
-\Lambda H_Y\Lambda^\top.
\end{aligned}
\end{equation}
The derivative formula for a normalized eigenvector gives $Dw[H]=RDM(\Sigma)[H]e$, while $Dr_c[H]=-H_Xc/2$. Hence
\begin{equation}\label{eq:coordinate-linear-expansion}
 D\theta_c(\Sigma)[H]
 =-\frac12c^\top H_Xe+c^\top RDM(\Sigma)[H]e.
\end{equation}
Expanding $R$ by \eqref{eq:reduced-resolvent-spectrum} gives, with $c_\ell=c^\top e_\ell$,
\begin{equation}\label{eq:coordinate-linear-expansion-modal}
\begin{aligned}
 D\theta_c(\Sigma)[H]
 ={}&-\frac{c_1}{2}H_{X,11}+\sum_{\ell=2}^p\frac{c_\ell}{\lambda_1^2-\lambda_\ell^2}
 \{-\lambda_1^2H_{X,\ell1}
   +\lambda_1H_{XY,\ell1}
   +\lambda_\ell H_{XY,1\ell}
   -\lambda_1\lambda_\ell H_{Y,\ell1}\},
\end{aligned}
\end{equation}
where the terms involving $\lambda_\ell$ are zero for $\ell>m$. For $H=ZZ^\top-\Sigma$, write
$\xi_\ell=e_\ell^\top X$ and $\eta_\ell=e_\ell^\top Y$. Then
\begin{equation}\label{eq:g-coordinate-score}
\begin{aligned}
 &g_c(X,Y)=-\frac{c_1}{2}(\xi_1^2-1)
             +\sum_{\ell=2}^p c_\ell g_\ell,\\
 &g_\ell=\frac{\lambda_1\xi_\ell(\eta_1-\lambda_1\xi_1)
       +\lambda_\ell\eta_\ell(\xi_1-\lambda_1\eta_1)}
      {\lambda_1^2-\lambda_\ell^2}.
\end{aligned}
\end{equation}
Again, the second numerator term is absent for $\ell>m$. For $2\le\ell\le m$, the component $g_\ell$ satisfies the conditions
of Lemma~\ref{lem:two-component-augmentation} with
\begin{equation}\label{eq:nonleading-component-parameters}
\begin{aligned}
 \phi&=\xi_1\xi_\ell,& \psi&=\eta_1\eta_\ell,&
 \rho&=\lambda_1\lambda_\ell,\\
 b_X&=\frac{(1-\lambda_1^2)\lambda_\ell^2}
                 {\lambda_1^2-\lambda_\ell^2},&
 b_Y&=\frac{(1-\lambda_1^2)\lambda_1\lambda_\ell}
                 {\lambda_1^2-\lambda_\ell^2}.
\end{aligned}
\end{equation}
For the leading component $g=-(\xi_1^2-1)/2$, use
\begin{equation}\label{eq:normalization-component-parameters}
\begin{gathered}
 \phi=(\xi_1^2-1)/\sqrt2,\qquad
 \psi=(\eta_1^2-1)/\sqrt2,\qquad \rho=\lambda_1^2,\\
 (b_X,b_Y)=-2^{-1/2}(1,\lambda_1^2),\qquad \Var(g)=1/2.
\end{gathered}
\end{equation}
Gaussian regression gives the conditional identities required by the lemma. Its coefficients reproduce the spectral coefficients of $G_X,G_Y$ in \eqref{eq:reduced-augmentation-spectral}. Therefore the
population optimal augmentation functions are
\begin{equation}\label{eq:coordinate-marginal-functions-general}
 a_c(X)=\langle B_{X,c}(\Sigma),XX^\top-I_p\rangle_{\Frob},\qquad
 b_c(Y)=\langle B_{Y,c}(\Sigma),YY^\top-I_q\rangle_{\Frob}.
\end{equation}
For $c=c_j$, expanding the two matrices in the canonical basis gives the functions in \eqref{eq:marginal-functions-main}, expressed in the original coordinates. Thus the reduced-resolvent representation recovers both the optimal augmentations and the implemented estimator without introducing coordinate-specific auxiliary vectors.

\subsubsection{Explicit augmentation functions}
\label{sec:explicit-augmentations}
We now give the Gaussian conditional-moment calculation and the explicit augmentation functions used in Section~\ref{sec:coordinate-method}, expressed in the original coordinates.
First, we define the population canonical variates
\[
    \xi_\ell = (u_\ell^*)^\top X,
    \qquad
    \eta_\ell = (v_\ell^*)^\top Y,
    \qquad 1 \le \ell \le m.
\]
The CCA normalization and orthogonality conditions give
\[
\begin{aligned}
    \mathbb{E}[\xi_\ell \xi_r]
    = \mathbb{E}[\eta_\ell \eta_r]
     = \mathbf{1}\{\ell=r\}\qquad \text{and} \qquad
    \mathbb{E}[\xi_\ell \eta_r]
    = \lambda_\ell \mathbf{1}\{\ell=r\}.
\end{aligned}
\]
Under joint Gaussianity, the CCA equations also imply
\[
    \mathbb{E}[\eta_\ell \mid X] = \lambda_\ell \xi_\ell,
    \qquad \text{and} \qquad
    \mathbb{E}[\xi_\ell \mid Y] = \lambda_\ell \eta_\ell.
\]
This structure extends to the quadratic terms. In particular, since 
\[
\operatorname{Var}(\eta_1\mid X) = \operatorname{Var}(\xi_1\mid Y) = 1-\lambda_1^2,
\]
we have
\[
\begin{aligned}
    \mathbb{E}[\eta_1^2-1 \mid X]
    &= \lambda_1^2(\xi_1^2-1)\qquad \text{and} \qquad
    \mathbb{E}[\xi_1^2-1 \mid Y]
    &= \lambda_1^2(\eta_1^2-1).
\end{aligned}
\]
For $2 \le \ell \le m$, the conditional covariances between distinct canonical variates are zero, so
\[
\begin{aligned}
    \mathbb{E}[\eta_1\eta_\ell \mid X]
    &= \lambda_1\lambda_\ell\,\xi_1\xi_\ell \qquad \text{and} \qquad    \mathbb{E}[\xi_1\xi_\ell \mid Y]
    &= \lambda_1\lambda_\ell\,\eta_1\eta_\ell.
\end{aligned}
\]
For the influence function $g_j$, its conditional expectations given $X$ and $Y$ are linear combinations of
\[
\{\xi_1^2-1,\ \xi_1\xi_2, \dots, \xi_1\xi_m\}
\text{ and }
\{\eta_1^2-1,\ \eta_1\eta_2, \dots, \eta_1\eta_m\},
\]
respectively. We solve the projection equations by expressing $a_j(X)$ and $b_j(Y)$ as linear combinations of these quadratic functions. For $2\le\ell\le m$, define
\[
\mathcal D_{\ell,\gamma} = (\lambda_1^2-\lambda_\ell^2) (1-\gamma_X\gamma_Y\lambda_1^2\lambda_\ell^2).
\]
Then the optimal augmentation functions are given by
\begin{equation}\label{eq:marginal-functions-main}
\begin{aligned}
a_j(X) =&-\frac{(e_j^\top u_1^*)(1-\gamma_Y\lambda_1^4)}{2(1-\gamma_X\gamma_Y\lambda_1^4)}(\xi_1^2-1) +\sum_{\ell=2}^m\frac{(e_j^\top u_\ell^*)(1-\lambda_1^2)\lambda_\ell^2(1-\gamma_Y\lambda_1^2)}{\mathcal D_{\ell,\gamma}}\xi_1\xi_\ell,\\
b_j(Y)=&-\frac{(e_j^\top u_1^*)\lambda_1^2(1-\gamma_X)}{2(1-\gamma_X\gamma_Y\lambda_1^4)}(\eta_1^2-1)+\sum_{\ell=2}^m\frac{(e_j^\top u_\ell^*)(1-\lambda_1^2)\lambda_1\lambda_\ell(1-\gamma_X\lambda_\ell^2)}{\mathcal D_{\ell,\gamma}}\eta_1\eta_\ell .
\end{aligned}
\end{equation}

Substituting the paired CCA estimates into \eqref{eq:marginal-functions-main} and taking pooled-minus-paired averages gives the corrections in \eqref{eq:shared-coordinate-correction}. The centering constants cancel in these differences, and the quadratic terms depend on the data only through $D_X$ and $D_Y$. This yields the implementation in \eqref{eq:ss-estimator}.

\subsection{Local smoothness and uniform derivative bounds}
\begin{lemma}[Uniform local derivatives]\label{lem:local-expansions}
Under Assumption~\ref{ass:model}, there is a fixed $r_0>0$ such that on $\|S-\Sigma\|_{\op}<r_0$ the maps $A^{-1/2},B^{-1/2},K,M,a,w,P,R$ are smooth, with the locally oriented leading eigenvector $w$. The maps $u,v,\lambda$ are three times continuously differentiable and, for $k=1,2,3$,
\begin{equation}\label{eq:local-derivative-bounds}
\begin{aligned}
 &\|D^ku(S)[H_1,\ldots,H_k]\|
 +\|D^kv(S)[H_1,\ldots,H_k]\|+|D^k\lambda(S)[H_1,\ldots,H_k]|
 \le C_k\prod_{r=1}^k\|H_r\|_{\op}.
\end{aligned}
\end{equation}
For every fixed bounded target $c$ and uniformly in $\gamma\in[0,1]^2$, the matrices $B_{X,c},B_{Y,c}$ in
\eqref{eq:coordinate-augmentation-matrices} satisfy
\begin{equation}\label{eq:local-matrix-coefficient-bounds}
 \|B_{W,c}(S)\|_*\le C\|c\|,
 \qquad
 \|DB_{W,c}(S)[H]\|_*\le C\|c\|\,\|H\|_{\op},\quad W=X,Y,
\end{equation}
where $\|\cdot\|_*$ is the nuclear norm. Analogous bounds, without the factor $\|c\|$, hold for $k_X\{\lambda(S)\}u(S)u(S)^\top$ and $k_Y\{\lambda(S)\}v(S)v(S)^\top$, where $k_X,k_Y$ are defined in
Appendix~\ref{sec:lambda-method}.
Moreover, for $T_c(S)=\theta_c(S)$ and $T_\lambda(S)=\lambda(S)$,
\begin{equation}\label{eq:target-derivative-lipschitz}
 |\{DT(S')-DT(S)\}[H]|
 \le C\|S'-S\|_{\op}\|H\|_{\op},
\end{equation}
and
\begin{equation}\label{eq:all-correlations-lipschitz}
 \max_{\ell\le m}|\lambda_\ell^2(S')-\lambda_\ell^2(S)|
 \le C\|S'-S\|_{\op}.
\end{equation}
In particular, every listed map is locally Lipschitz. We use
\begin{equation}\label{eq:direct-lipschitz}
 \|u(S')-u(S)\|+\|v(S')-v(S)\|+|\lambda(S')-\lambda(S)|
 \le C\|S'-S\|_{\op}.
\end{equation}
\end{lemma}

\begin{proof}
Uniform positive definiteness of $A,B$ makes inversion and inverse square roots smooth with fixed-order derivative bounds. Hence $K$ and $M$ are smooth, with derivatives bounded by products of operator norms of the increments. Weyl's inequality gives \eqref{eq:all-correlations-lipschitz} and preserves the leading gap on a ball of fixed radius. Consider the symmetric eigenpair equations
\[
 Mw=aw,\qquad w^\top w=1.
\]
On $w^\perp$, the inverse of $aI-M$ has operator norm at most $\{a(S)-\lambda_2^2(S)\}^{-1}$, which is uniformly bounded by the eigen-gap assumption. The first derivative is
\begin{equation}\label{eq:reduced-eigen-derivative}
 Da(S)[H]=w^\top DM(S)[H]w,
 \qquad
 Dw(S)[H]=R(S)DM(S)[H]w.
\end{equation}
Differentiating the eigenpair equations two and three times, the highest order derivative again enters through this same invertible operator. The right side contains only derivatives of $M$ and lower-order derivatives of $(a,w)$. Induction therefore gives dimension-free bounds for $D^ka,D^kw$, $k\le3$. Since
\[
 \lambda=a^{1/2},\qquad
 u=A^{-1/2}w,\qquad
 v=\lambda^{-1}B^{-1/2}K^\top w,
\]
and $\lambda$ stays bounded away from zero, this proves \eqref{eq:local-derivative-bounds}.

The matrices $P,R,J,G_X,G_Y$ are obtained from $M,a,w$ by a fixed number of matrix products and inversions. The matrices being inverted have eigenvalues uniformly bounded away from zero. The inverse identity $X^{-1}-Y^{-1}=X^{-1}(Y-X)Y^{-1}$ and differentiation therefore give uniform first-derivative bounds for these maps. Each matrix in \eqref{eq:coordinate-augmentation-matrices} is the symmetrization of a single outer product with uniformly bounded factors. Its derivative is a fixed sum of such outer products, which proves \eqref{eq:local-matrix-coefficient-bounds}. The same argument applies to the two rank-one correlation coefficients. Finally, $\theta_c=r_c^\top w$ with $r_c=A^{-1/2}c$, so the second-derivative bound implies \eqref{eq:target-derivative-lipschitz}. Integrating the first derivative bounds along line segments gives \eqref{eq:direct-lipschitz}.
\end{proof}

\subsection{General expansion for augmented estimators}
\label{sec:stochastic-bounds}

\begin{lemma}[First-order expansion of an augmented estimator]
\label{lem:augmented-expansion}
Under the Gaussian sampling model and Assumption~\ref{ass:model}, suppose $d=o(n)$. Let $\theta$ be a continuously differentiable real-valued function on a neighborhood of $\Sigma$ such that
\[
 |D\theta(S)[H]|\le C\|H\|_{\op},\qquad
 |\{D\theta(S')-D\theta(S)\}[H]|
 \le C\|S'-S\|_{\op}\|H\|_{\op}.
\]
Let $B_X(S),B_Y(S)$ be symmetric matrix maps satisfying
\begin{equation}\label{eq:generic-matrix-coefficient-bounds}
 \|B_W(S)\|_*\le C,
 \qquad
 \|DB_W(S)[H]\|_*\le C\|H\|_{\op},\quad W=X,Y.
\end{equation}
Choose a fixed $\delta>0$ so that $\{\|S-\Sigma\|_{\op}<2\delta\}$ lies in the common neighborhood and $\delta<\lambda_{\min}(\Sigma)$, and let $\mathcal E_n$ be the event in \eqref{eq:local-good-event}. Consider, on $\mathcal E_n$,
\begin{equation}\label{eq:generic-augmented-estimator}
 \widehat\theta=\theta(\widehat\Sigma)
 +\langle B_X(\widehat\Sigma),D_X\rangle_{\Frob}
 +\langle B_Y(\widehat\Sigma),D_Y\rangle_{\Frob}
 -\widehat{\mathcal C}_n.
\end{equation}
For $\|E\|_{\op}<2\delta$, define
\begin{equation}\label{eq:generic-paired-functional}
\begin{aligned}
 F(E)&=\theta(\Sigma+E)
 -\gamma_X\langle B_X(\Sigma+E),E_X\rangle_{\Frob}
 -\gamma_Y\langle B_Y(\Sigma+E),E_Y\rangle_{\Frob},\\
 R_F(E)&=F(E)-F(0)-DF(0)[E],\qquad
 \mu_{F,n}=\E\{R_F(\Delta)\mid\mathcal E_n\}.
\end{aligned}
\end{equation}
For $W=X,Y$, write $B_W=B_W(\Sigma)$ and $\widehat B_W=B_W(\widehat\Sigma)$ for the population and fitted coefficient matrices, respectively. Set
\begin{equation}\label{eq:generic-linear-functions}
\begin{aligned}
 g_\theta(Z)&=D\theta(\Sigma)[ZZ^\top-\Sigma],\\
 a_\theta(X)&=\langle B_X,XX^\top-\Sigma_X\rangle_{\Frob},\\
 b_\theta(Y)&=\langle B_Y,YY^\top-\Sigma_Y\rangle_{\Frob},
\end{aligned}
\end{equation}
and
\begin{equation}\label{eq:generic-linear-average}
\begin{aligned}
 \mathcal L_{\theta,n}={}&\frac1n\sum_{i=1}^n
 \{g_\theta(Z_i)-\gamma_Xa_\theta(X_i)-\gamma_Yb_\theta(Y_i)\}+\frac1{n+N_X}\sum_{r=1}^{N_X}a_\theta(X_r^U)
 +\frac1{n+N_Y}\sum_{s=1}^{N_Y}b_\theta(Y_s^U).
\end{aligned}
\end{equation}
Then, 
\begin{equation}\label{eq:generic-one-step-expansion}
 \widehat\theta-\theta(\Sigma)
 =\mathcal L_{\theta,n}+\mu_{F,n}-\widehat{\mathcal C}_n
 +O_p(\sqrt d/n).
\end{equation}
If an orthogonal $T$ satisfies $T\Sigma T^\top=\Sigma$ and
\begin{equation}\label{eq:remainder-reflection-condition}
 F(TET^\top)=-F(E)\qquad(\|E\|_{\op}<2\delta),
\end{equation}
then $\mu_{F,n}=0$. Alternatively, if a deterministic homogeneous quadratic polynomial $H$ satisfies $|H(E)|\le C\|E\|_{\op}^2$ and $R_F(E)=H(E)+O(\|E\|_{\op}^3)$ on the neighborhood, then
\begin{equation}\label{eq:remainder-conditional-mean}
 \mu_{F,n}=n^{-1}\E H(\Psi)+O((d/n)^{3/2}).
\end{equation}
\end{lemma}

\begin{proof}
For a coefficient map $B(S)$ and $S=\Sigma+E$,
\[
 D\langle B(S),E_W\rangle_{\Frob}[H]
 =\langle B(S),H_W\rangle_{\Frob}
   +\langle DB(S)[H],E_W\rangle_{\Frob}.
\]
The nuclear/operator duality and \eqref{eq:generic-matrix-coefficient-bounds} give
\begin{equation}\label{eq:augmented-derivative-lipschitz}
 |\{DF(E)-DF(0)\}[H]|
 \le C\|E\|_{\op}\|H\|_{\op}.
\end{equation}
Therefore
\begin{equation}\label{eq:augmented-gradient-bound}
 \|\nabla R_F(E)\|_{\Frob}
 \le C\|E\|_{\op},
 \qquad
 |R_F(E)|\le C\|E\|_{\op}^2.
\end{equation}
The covariance concentration bound gives, for a sufficiently small fixed $c_1>0$ and all large $n$,
\begin{equation}\label{eq:local-event-tail}
 \Pr(\mathcal E_n^c)\le e^{-c_1n},\qquad
 \Pr(\mathcal E_n)\ge\tfrac12,\qquad n-d-1\ge n/2.
\end{equation}
Applying Lemma~\ref{lem:conditional-wishart} to $h(S)=R_F(S-\Sigma)$ gives
\begin{equation}\label{eq:remainder-conditional-variance}
 \Var\{R_F(\Delta)\mid\mathcal E_n\}
 \le \frac Cn\E\{\|\Delta\|_{\op}^2\mid\mathcal E_n\}
 \le Cd/n^2.
\end{equation}
Hence $R_F(\Delta)-\mu_{F,n}=O_p(\sqrt d/n)$. Using $D_W=U_W-\gamma_W\Delta_W$, on $\mathcal E_n$ we have
\[
 \widehat\theta-\theta(\Sigma)
 =DF(0)[\Delta]+R_F(\Delta)
 +\langle \widehat B_X,U_X\rangle_{\Frob}
 +\langle \widehat B_Y,U_Y\rangle_{\Frob}
 -\widehat{\mathcal C}_n.
\]
The fitted-coefficient error is
\begin{equation}\label{eq:generic-unpaired-difference}
 \langle \widehat B_W-B_W,U_W\rangle_{\Frob}.
\end{equation}
On $\mathcal E_n$, $\|\widehat B_W-B_W\|_{\Frob}\le \|\widehat B_W-B_W\|_*\le C\|\Delta\|_{\op}$. Equation~\eqref{eq:localized-mixed-second-moment} therefore implies that the term in \eqref{eq:generic-unpaired-difference} is $O_p(\sqrt d/n)$. Moreover,
\[
 DF(0)[E]=D\theta(\Sigma)[E]
 -\gamma_X\langle B_X,E_X\rangle_{\Frob}
 -\gamma_Y\langle B_Y,E_Y\rangle_{\Frob},
\]
so the linear terms with coefficients evaluated at the population covariance combine to give $\mathcal L_{\theta,n}$. This proves \eqref{eq:generic-one-step-expansion}. For the reflection case, differentiation of \eqref{eq:remainder-reflection-condition} at zero gives
\[
DF(0)[TET^\top]=-DF(0)[E],
\] 
so $R_F(TET^\top)=-R_F(E)$. Both the Gaussian law and the event are reflection invariant. Hence the conditional expectation equals its negative and $\mu_{F,n}=0$.

For the case with quadratic approximation $H$, homogeneity and independence give $\E H(\Delta)=n^{-1}\E H(\Psi)$. The difference between this mean and $\E\{H(\Delta)\mid\mathcal E_n\}$ is exponentially small by Cauchy-Schwarz, \eqref{eq:cov-conc}, and
\eqref{eq:local-event-tail}. Finally,
\[
 |\E\{R_F(\Delta)-H(\Delta)\mid\mathcal E_n\}|
 \le C\frac{\E\|\Delta\|_{\op}^3}{\Pr(\mathcal E_n)}
 \le C(d/n)^{3/2},
\]
which proves \eqref{eq:remainder-conditional-mean}.
\end{proof}

\section{Inference for leading canonical directions}\label{sec:coordinate-null-supp}
\subsection{Asymptotic variance and its estimation}
\label{sec:coordinate-variance-proof}

For the original coordinate $j$, the fixed target is $c_j$ in \eqref{eq:transformed-test-functional}. Under the null $u_{1j}^*=0$, its first coordinate is zero. Hence
\eqref{eq:g-coordinate-score} reduces to
\begin{equation}\label{eq:g-canonical-direct}
 g_{c_j}(X,Y)=\sum_{\ell=2}^p(e_j^\top u_\ell^*)g_\ell(X,Y).
\end{equation}
Here $g_{c_j}$ and $g_j$ denote the same influence function in canonical and original coordinates, respectively. Let $g_\ell$ denote the $\ell$-th component in \eqref{eq:g-coordinate-score}. Conditional on the leading pair $(\xi_1,\eta_1)$, distinct non-leading components are independent and centered, so their covariance is zero. For $2\le\ell\le m$, Isserlis' identity gives
\[
\begin{aligned}
 \Var\{\lambda_1\xi_\ell(\eta_1-\lambda_1\xi_1)        +\lambda_\ell\eta_\ell(\xi_1-\lambda_1\eta_1)\} =(1-\lambda_1^2)
       (\lambda_1^2+\lambda_\ell^2-2\lambda_1^2\lambda_\ell^2),
\end{aligned}
\]
because
\[
\begin{aligned}
 &\E\{\xi_\ell^2(\eta_1-\lambda_1\xi_1)^2\}=1-\lambda_1^2,\\
 &\E\{\eta_\ell^2(\xi_1-\lambda_1\eta_1)^2\}=1-\lambda_1^2,\\
 &\E\{\xi_\ell\eta_\ell(\eta_1-\lambda_1\xi_1)
                     (\xi_1-\lambda_1\eta_1)\}=-\lambda_1\lambda_\ell(1-\lambda_1^2).
\end{aligned}
\]
Dividing by $(\lambda_1^2-\lambda_\ell^2)^2$ gives $\tau_{0,\ell}^2$. Applying Lemma~\ref{lem:two-component-augmentation} with \eqref{eq:nonleading-component-parameters} gives
\[
 \tau_{0,\ell}^2-\tau_{\gamma,\ell}^2
 =\frac{\gamma_Xb_X^2+\gamma_Yb_Y^2
        -2\gamma_X\gamma_Y(\lambda_1\lambda_\ell)b_Xb_Y}
       {1-\gamma_X\gamma_Y\lambda_1^2\lambda_\ell^2},
\]
which simplifies to \eqref{eq:modal-gain-main}. For $m<\ell\le p$, $g_\ell=\xi_\ell(\eta_1-\lambda_1\xi_1)/\lambda_1$. Its variance is $(1-\lambda_1^2)/\lambda_1^2$, and both marginal projections vanish. This is the same formula with $\lambda_\ell=0$. Summing the uncorrelated components gives \eqref{eq:sigma-two-sided}.

\begin{proof}[Proof of Proposition~\ref{prop:var-reduction}]
The preceding calculation gives
\[
 \sigma_{j,n}^2(0,0)-\sigma_{j,n}^2(\gamma_X,\gamma_Y)
 =\sum_{\ell=2}^p(e_j^\top u_\ell^*)^2\Delta_{\gamma,\ell}.
\]
Under Assumption~\ref{ass:model}, the denominator and the factors $1-\lambda_1^2$, $1-\gamma_Y\lambda_1^2$, and $1-\gamma_X\lambda_\ell^2$ in \eqref{eq:modal-gain-main} are positive. Thus $\Delta_{\gamma,\ell}>0$ exactly when $\gamma_X+\gamma_Y>0$ and $\lambda_\ell>0$, and it is zero otherwise. The sum is therefore positive exactly when $\gamma_X+\gamma_Y>0$ and $(e_j^\top u_\ell^*)\lambda_\ell\ne0$ for some $\ell\in\{2,\ldots,m\}$, which proves the proposition.
\end{proof}

\begin{lemma}[Coordinate variance]\label{lem:coordinate-variance}
Under Assumption~\ref{ass:model} and $u_{1j}^*=0$, $0<c_0\le\sigma_{j,n}^2\le C_0$ uniformly over $\gamma\in[0,1]^2$. If $d=o(n)$, then
\begin{equation}\label{eq:coordinate-variance-consistency}
 \widehat\sigma_j^2-\sigma_{j,n}^2=O_p(\sqrt{d/n}),\qquad
 \widehat\sigma_j^2/\sigma_{j,n}^2\xrightarrow{p}1.
\end{equation}
\end{lemma}

\begin{proof}
For $2\le\ell\le m$, let $\alpha_{\gamma,\ell}$ and $\beta_{\gamma,\ell}$ be the coefficients in \eqref{eq:two-component-coefficients}, evaluated at the parameters in \eqref{eq:nonleading-component-parameters}. The monotonicity statement in Lemma~\ref{lem:two-component-augmentation} gives
\begin{equation}\label{eq:variance-monotonicity}
 \partial_{\gamma_X}\tau_{\gamma,\ell}^2
 =-\alpha_{\gamma,\ell}^2,
 \qquad
 \partial_{\gamma_Y}\tau_{\gamma,\ell}^2
 =-\beta_{\gamma,\ell}^2.
\end{equation}
For $\ell>m$, the component variances do not depend on $\gamma$. Consequently,
\[
 \tau_{(1,1),\ell}^2\le\tau_{\gamma,\ell}^2\le\tau_{0,\ell}^2,
 \qquad
 \tau_{(1,1),\ell}^2
 =\tau_{0,\ell}^2\frac{1-\lambda_\ell^2}
                             {1-\lambda_1^2\lambda_\ell^2}.
\]
The endpoints are uniformly bounded above and below from zero. Since $c_{j1}=0$ and
\[
 \sum_{\ell=2}^p c_{j\ell}^2
 =\|c_j\|^2=e_j^\top\Sigma_X^{-1}e_j\asymp1,
\]
where $\Sigma_X$ in this display denotes the original marginal covariance used to define $c_j$, we obtain
\begin{equation}\label{eq:sigma-bounds}
 0<c_0\le\sigma_{j,n}^2\le C_0.
\end{equation}

It remains to control the plug-in variance without choosing non-leading eigenvectors. For a sample covariance $S$, use $a,M,R,J,r_c$ from \eqref{eq:one-sided-cca-operator}-\eqref{eq:reduced-target}, with $c=c_j$, and put $\eta=\gamma_X\gamma_Y$. Here the matrices and canonical correlations are evaluated at $S$. Define the matrix polynomial
\begin{equation}\label{eq:modal-variance-polynomial}
\begin{aligned}
 N_\gamma(a,M)={}&(1-a)(aI+M-2aM)(I-\eta aM)\\
 &-(1-a)^2\{\gamma_X(1-\gamma_Ya)M^2
        +\gamma_YaM(I-\gamma_XM)\}.
\end{aligned}
\end{equation}
Then set
\begin{equation}\label{eq:coordinate-matrix-variance}
 \mathcal V_c(S)=r_c^\top N_\gamma(a,M)R^2Jr_c.
\end{equation}
In an eigenbasis of $M$, $Rw_1=0$. For each non-leading eigenvalue $x=\lambda_\ell^2(S)$, the corresponding scalar coefficient is
\[
\begin{aligned}
 \tau_{\gamma,\ell}^2
 ={}&\frac{(1-a)(a+x-2ax)}{(a-x)^2}-\frac{(1-a)^2\{\gamma_Xx^2(1-\gamma_Ya)
                 +\gamma_Yax(1-\gamma_Xx)\}}
 {(a-x)^2(1-\eta ax)}.
\end{aligned}
\]
Consequently,
\begin{equation}\label{eq:coordinate-matrix-variance-identities}
 \mathcal V_{c_j}(\Sigma)=\sigma_{j,n}^2,
 \qquad
 \mathcal V_{c_j}(\widehat\Sigma)=\widehat\sigma_j^2
\end{equation}
whenever the $S$ lies in the neighborhood of $\Sigma$. This identity is invariant under rotations inside repeated non-leading eigenspaces. 

Lemma~\ref{lem:local-expansions}, the inverse identity, and the polynomial form \eqref{eq:modal-variance-polynomial} give
\begin{equation}\label{eq:local-null-variance-bound}
 |\mathcal V_{c_j}(S)-\mathcal V_{c_j}(\Sigma)|
 \le C\|S-\Sigma\|_{\op}.
\end{equation}
Thus \eqref{eq:cov-conc} proves $\widehat\sigma_j^2-\sigma_{j,n}^2=O_p(\sqrt{d/n})$. Together with \eqref{eq:sigma-bounds}, this proves consistency. Finally, choose the local radius so that the right side of \eqref{eq:local-null-variance-bound} is at most $c_0/2$. Then $\mathcal V_{c_j}(S)\ge c_0/2$ throughout the neighborhood.
\end{proof}

\subsection{Proof of Theorem~\ref{thm:main}}\label{sec:coordinate-theorem-proof}
\begin{proof}
Work in the canonical coordinates of \eqref{eq:fixed-canonical-transform}, where $c_{j1}=0$. Applying Lemma~\ref{lem:augmented-expansion} with
\[
 \theta(S)=\theta_{c_j}(S),\qquad
 B_X(S)=B_{X,c_j}(S),\qquad
 B_Y(S)=B_{Y,c_j}(S),\qquad
 \widehat{\mathcal C}_n=0,
\]
gives
\[
\widetilde u_{1j}^{\mathrm{Semi}}-u_{1j}^* = \mathcal L_{j,n} +\mu_{F_{c_j},n} +O_p(\sqrt d/n).
\]
Lemma~\ref{lem:local-expansions} verifies the required regularity conditions, while \eqref{eq:ss-matrix-implementation} shows that the resulting estimator equals the signed semi-supervised coordinate estimator on $\mathcal E_n$. By Lemma~\ref{lem:coordinate-variance}, we may choose the neighborhood of $\Sigma$ sufficiently small so that $\widehat\sigma_j^2$ is bounded away from zero on $\mathcal E_n$.

To show that the conditional mean of the nonlinear remainder vanishes under the null, consider the reflection that preserves the first canonical coordinate and reverses the signs of all remaining coordinates in both views
\[
 T_X=\diag(1,-I_{p-1}),\qquad
 T_Y=\diag(1,-I_{q-1}),\qquad
 T=\diag(T_X,T_Y).
\]
The reflection preserves the population covariance since $T_X\Lambda T_Y=\Lambda$. Moreover, $T_Xc_j=-c_j$
under the null. For $S'=TST^\top$ in the local ball, orthogonal functional calculus gives
\[
\begin{aligned}
 K'&=T_XKT_Y,& M'&=T_XMT_X,& w'&=T_Xw,\\
 P'&=T_XPT_X,& R'&=T_XRT_X,& J'&=T_XJT_X.
\end{aligned}
\]
The sign convention is preserved because $T_Xe=e$. Moreover, $r_{c_j}(S')=-T_Xr_{c_j}(S)$, and hence
\begin{equation}\label{eq:coordinate-reflection-matrices}
 B_{X,c_j}(S')=-T_XB_{X,c_j}(S)T_X,
 \qquad
 B_{Y,c_j}(S')=-T_YB_{Y,c_j}(S)T_Y.
\end{equation}
Therefore the paired functional in
\eqref{eq:generic-paired-functional} satisfies
\begin{equation}\label{eq:coordinate-reflection-neighborhood}
 F_{c_j}(TET^\top)=-F_{c_j}(E)
 \qquad (\|E\|_{\op}<2\delta).
\end{equation}
Since $Z_i\sim N(0,\Sigma)$ and $T\Sigma T^\top=\Sigma$, the reflected observations $TZ_i$ have the same distribution as $Z_i$. Moreover, $\|T\Delta T^\top\|_{\op}=\|\Delta\|_{\op}$, so the event $\mathcal E_n$ is preserved. Together with $R_{F_{c_j}}(T\Delta T^\top)=-R_{F_{c_j}}(\Delta)$, this implies $\mu_{F_{c_j},n}=0$.

For $\theta=\theta_{c_j}$, the functions in \eqref{eq:generic-linear-functions} coincide with $g_j,a_j,b_j$ when expressed in the original coordinates. Therefore, the first-order term in \eqref{eq:generic-linear-average} is
\begin{equation}\label{eq:coordinate-linear-averages}
\begin{aligned}
 \mathcal L_{j,n}={}&\frac1n\sum_{i=1}^n
 \{g_j(X_i,Y_i)-\gamma_Xa_j(X_i)-\gamma_Yb_j(Y_i)\}+\frac1{n+N_X}\sum_{r=1}^{N_X}a_j(X_r^U)
  +\frac1{n+N_Y}\sum_{s=1}^{N_Y}b_j(Y_s^U).
\end{aligned}
\end{equation}
The expansion yields, under the original sampling law,
\begin{equation}\label{eq:coordinate-asymptotic-linear}
 \widetilde u_{1j}^{\rm Semi}-u_{1j}^* =\mathcal L_{j,n}+O_p(\sqrt d/n).
\end{equation}
Independence of the three sample groups and $nN_W/(n+N_W)^2=\gamma_W(1-\gamma_W)$ give

\begin{equation*}
 \Var(\sqrt n\,\mathcal L_{j,n})
 =\Var(g_j-\gamma_Xa_j-\gamma_Yb_j) +\gamma_X(1-\gamma_X)\Var(a_j) +\gamma_Y(1-\gamma_Y)\Var(b_j)
 =\sigma_{j,n}^2.
\end{equation*}

The summands are centered Gaussian quadratic forms with uniformly bounded variances, so \eqref{eq:gaussian-fourth-moment}, \eqref{eq:common-lyapunov-bound}, and \eqref{eq:sigma-bounds} verify Lyapunov's condition. Since $d=o(n)$, the remainder in \eqref{eq:coordinate-asymptotic-linear} is $o_p(n^{-1/2})$. Lemma~\ref{lem:coordinate-variance} and \eqref{eq:prelim-studentization} therefore give
\[
 \frac{n(\widetilde u_{1j}^{\rm Semi})^2}{\widehat\sigma_j^2}
 \rightsquigarrow\chi_1^2.
\]
Changing the sign of the leading canonical directions also changes the sign of the augmentation matrices in \eqref{eq:coordinate-augmentation-matrices}. Thus, the semi-supervised coordinate estimator changes sign, while its variance estimator remains unchanged. Moreover, the matrix representation is invariant to the choice of non-leading eigenvectors, even when their eigenvalues coincide. Therefore, the implemented statistic $W_j$ equals the displayed statistic on $\mathcal E_n$. Since $\Pr(\mathcal E_n^c)\to0$ and the chi-squared distribution is continuous, the test has asymptotic size $\alpha$.\end{proof}

\section{Inference for leading canonical correlation}
\label{sec:lambda-method}\label{sec:lambda-main}

Define
\[
 k_X(\lambda)=\frac{\lambda(1-\lambda^2)(1-\gamma_Y\lambda^2)}
                    {2(1-\gamma_X\gamma_Y\lambda^4)},\qquad
 k_Y(\lambda)=k_X(\lambda)\big|_{\gamma_X\leftrightarrow\gamma_Y}.
\]
With the explicit correction $\widehat{\mathcal C}_{\lambda,n}$ in
\eqref{eq:C-lambda-hat}, the estimator is
\begin{equation}\label{eq:lambda-ss-estimator}
 \widehat\lambda_1^{\mathrm{Semi}}=\widehat\lambda_1
 +k_X(\widehat\lambda_1)\widehat u_1^\top D_X\widehat u_1
 +k_Y(\widehat\lambda_1)\widehat v_1^\top D_Y\widehat v_1
 -\widehat{\mathcal C}_{\lambda,n}.
\end{equation}
Write $s_\gamma^2(\lambda)$ for the variance function
\begin{equation}\label{eq:sigma-lambda-n}
 \sigma_{\lambda,n}^2=s_\gamma^2(\lambda_1)
 =(1-\lambda_1^2)^2
 \frac{1-\tfrac12(\gamma_X+\gamma_Y)\lambda_1^2}
      {1-\gamma_X\gamma_Y\lambda_1^4},
\end{equation}
and set
{\begin{equation}\label{eq:sigma-lambda-hat}
 \widehat\sigma_\lambda^2=s_\gamma^2(\widehat\lambda_1).
\end{equation}}

Unlike in Section~\ref{sec:coordinate-null-supp}, we do not impose a null hypothesis for inference for the leading canonical correlation. The fluctuations of the nonlinear remainder remain $O_p(\sqrt d/n)$, but its mean generally does not vanish. We therefore derive an analytic correction for the second-order bias.
\subsection{Influence function and variance}\label{sec:lambda-influence-proof}

Let $(X^\top,Y^\top)^\top\sim N(0,S)$, where $S=\left(\begin{smallmatrix}A&C\\C^\top&B\end{smallmatrix}\right)$, and let $(u,v,\lambda)$ be its normalized leading canonical triplet. Expectations and variances in this calculation are taken under $N(0,S)$. Equation~\eqref{eq:lambda-linear-expansion} gives
\begin{equation}\label{eq:g-lambda}
 g_{\lambda,S}(X,Y)=(u^\top X)(v^\top Y)
       -\frac\lambda2\{(u^\top X)^2+(v^\top Y)^2\}.
\end{equation}
Write $\xi=u^\top X$, $\eta=v^\top Y$. Gaussian conditioning gives
\[
 \E(g_{\lambda,S}\mid X)=\tfrac12\lambda(1-\lambda^2)(\xi^2-1),
 \qquad
 \E(g_{\lambda,S}\mid Y)=\tfrac12\lambda(1-\lambda^2)(\eta^2-1).
\]
Apply Lemma~\ref{lem:two-component-augmentation} with
\begin{equation}\label{eq:lambda-component-parameters}
 \phi=\frac{\xi^2-1}{\sqrt2},\qquad
 \psi=\frac{\eta^2-1}{\sqrt2},\qquad
 \rho=\lambda^2,\qquad b_X=b_Y=\frac{\lambda(1-\lambda^2)}{\sqrt2}.
\end{equation}
The conditional-square identities give \eqref{eq:two-component-conditionals}. Its coefficients satisfy $\alpha_\gamma/\sqrt2=k_X(\lambda)$ and $\beta_\gamma/\sqrt2=k_Y(\lambda)$, so
\begin{equation}\label{eq:lambda-marginal-general}
 a_{\lambda,S}^\gamma(X)=k_X(\lambda)\{(u^\top X)^2-1\},\qquad
 b_{\lambda,S}^\gamma(Y)=k_Y(\lambda)\{(v^\top Y)^2-1\}.
\end{equation}
Equivalently, define the rank-one coefficient matrices
\begin{equation}\label{eq:lambda-augmentation-matrices}
 B_{\lambda,X}(S)=k_X\{\lambda(S)\}u(S)u(S)^\top,\qquad
 B_{\lambda,Y}(S)=k_Y\{\lambda(S)\}v(S)v(S)^\top.
\end{equation}
Then $a_{\lambda,S}^\gamma(X)=\langle B_{\lambda,X}(S),XX^\top-A\rangle_{\Frob}$ and similarly for $Y$. Write $g_\lambda,a_\lambda,b_\lambda$ for these functions evaluated at $S=\Sigma$. Their fitted pooled-minus-paired averages give exactly the two augmentation terms in \eqref{eq:lambda-ss-estimator}.

To evaluate the variance input to the same lemma, \eqref{eq:prelim-isserlis} gives $\E(\xi\eta)^2=1+2\lambda^2$, $\E\{\xi\eta(\xi^2+\eta^2)\}=6\lambda$, and $\E(\xi^2+\eta^2)^2=8+4\lambda^2$. Therefore
\[
 \Var(g_{\lambda,S})
 =1+2\lambda^2-6\lambda^2+\tfrac14\lambda^2(8+4\lambda^2)
 =(1-\lambda^2)^2.
\]
Substituting \eqref{eq:lambda-component-parameters} into \eqref{eq:two-component-variance} gives
\begin{equation}\label{eq:lambda-gain}
 (1-\lambda_1^2)^2-\sigma_{\lambda,n}^2
 =\frac{\lambda_1^2(1-\lambda_1^2)^2}{2}
   \frac{\gamma_X+\gamma_Y-2\gamma_X\gamma_Y\lambda_1^2}
        {1-\gamma_X\gamma_Y\lambda_1^4}.
\end{equation}
Subtracting this from the paired variance yields \eqref{eq:sigma-lambda-n}. The variance reduction is strictly positive whenever $\gamma_X+\gamma_Y>0$, even when the cross-covariance has rank one. Under Assumption~\ref{ass:model},
\begin{equation}\label{eq:sigma-lambda-bounds}
 0<c_0\le\sigma_{\lambda,n}^2\le C_0.
\end{equation}
\subsection{Second-order bias correction}\label{sec:analytic-implementation}
Set
\begin{equation}\label{eq:kappa-lambda}
 \kappa(\lambda)=\gamma_Xk_X(\lambda)+\gamma_Yk_Y(\lambda)
 =\frac{\lambda(1-\lambda^2)}2
    \frac{\gamma_X+\gamma_Y-2\gamma_X\gamma_Y\lambda^2}
         {1-\gamma_X\gamma_Y\lambda^4}.
\end{equation}
Its derivative is
\begin{equation}\label{eq:kappa-lambda-derivative}
\begin{aligned}
 \kappa'(\lambda)={}&
 \frac{(1-3\lambda^2)(\gamma_X+\gamma_Y-2\gamma_X\gamma_Y\lambda^2)}
      {2(1-\gamma_X\gamma_Y\lambda^4)}-\frac{2\gamma_X\gamma_Y\lambda^2(1-\lambda^2)
                    (1-\gamma_X\lambda^2)(1-\gamma_Y\lambda^2)}
             {(1-\gamma_X\gamma_Y\lambda^4)^2}.
\end{aligned}
\end{equation}
The correction used by the estimator is
\begin{equation}\label{eq:C-lambda-hat}
\begin{aligned}
 \widehat{\mathcal C}_{\lambda,n}=\frac1n\bigg[&
 \frac{1-\widehat\lambda_1^2}{2}
       \left\{\frac{p+q-2m}{\widehat\lambda_1}-\widehat\lambda_1\right\}
 +2\kappa(\widehat\lambda_1)
 -\widehat\lambda_1(1-\widehat\lambda_1^2)\kappa'(\widehat\lambda_1)\\
 &+(1-\widehat\lambda_1^2)\sum_{\ell=2}^m
 \frac{\widehat\lambda_1(1-\widehat\lambda_\ell^2)
                    -2\kappa(\widehat\lambda_1)\widehat\lambda_\ell^2}
      {\widehat\lambda_1^2-\widehat\lambda_\ell^2}\bigg].
\end{aligned}
\end{equation}
We derive the second-order bias using only the first derivatives of the canonical directions. Related higher-order CCA expansions have been studied by \citet{Ogasawara2007}. Our calculation also accounts for the estimation of the marginal augmentation functions.

\paragraph{Second-order Taylor expansion.}
We use the population canonical coordinates defined in \eqref{eq:fixed-canonical-transform}, with the transformed-quantity convention from Appendix~\ref{sec:coordinate-calculations}. Denote $\lambda=\lambda_1$, $e=e_1\in\R^p$, and $f=e_1\in\R^q$. For a symmetric increment $E$, set
\begin{equation}\label{eq:cca-first-increments}
\begin{aligned}
 \epsilon_X&=e^\top E_Xe,&\epsilon_Y&=f^\top E_Yf,\\
 L(E)&=e^\top E_{XY}f-\tfrac\lambda2(\epsilon_X+\epsilon_Y),\\
 \zeta_X(E)&=E_{XY}f-\lambda E_Xe,&
 \zeta_Y(E)&=E_{XY}^\top e-\lambda E_Yf.
\end{aligned}
\end{equation}
The solutions of \eqref{eq:cca-linear-equations} are
\begin{equation}\label{eq:cca-first-order-coordinates}
\begin{aligned}
 (h_X)_1&=-\epsilon_X/2,&(h_Y)_1&=-\epsilon_Y/2,\\
 (h_X)_\ell&=\frac{\lambda\zeta_{X,\ell}+\lambda_\ell\zeta_{Y,\ell}}
                         {\lambda^2-\lambda_\ell^2},&
 (h_Y)_\ell&=\frac{\lambda_\ell\zeta_{X,\ell}+\lambda\zeta_{Y,\ell}}
                         {\lambda^2-\lambda_\ell^2},\quad 2\le\ell\le m,\\
 (h_X)_\ell&=\zeta_{X,\ell}/\lambda\ (m<\ell\le p),&
 (h_Y)_\ell&=\zeta_{Y,\ell}/\lambda\ (m<\ell\le q).
\end{aligned}
\end{equation}
Along $S(t)=\Sigma+tE$, the exact first derivative is
\[
 \lambda'(t)=u(t)^\top E_{XY}v(t)
 -\frac{\lambda(t)}2\{u(t)^\top E_Xu(t)+v(t)^\top E_Yv(t)\}.
\]
Differentiation at zero, using $u(0)=e,v(0)=f$, gives
\[
\begin{aligned}
 \lambda''(0)
 ={}&h_X^\top E_{XY}f+e^\top E_{XY}h_Y
       -\tfrac12L(E)(\epsilon_X+\epsilon_Y)-\lambda\{h_X^\top E_Xe+h_Y^\top E_Yf\}\\
 ={}&\zeta_X^\top h_X+\zeta_Y^\top h_Y
       -\tfrac12L(E)(\epsilon_X+\epsilon_Y).
\end{aligned}
\]
Thus
\begin{equation}\label{eq:cca-correlation-quadratic}
 Q_\lambda(E)=\tfrac12\lambda''(0)
 =\tfrac12(\zeta_X^\top h_X+\zeta_Y^\top h_Y)
                    -\tfrac14L(E)(\epsilon_X+\epsilon_Y).
\end{equation}
The third-order bound in Lemma~\ref{lem:local-expansions} controls the remainder directly. Along $S(t)=\Sigma+tE$,
\[
 |\lambda'''(t)|
 =|D^3\lambda\{S(t)\}[E,E,E]|
 \le C\|E\|_{\op}^3,
 \qquad 0\le t\le1.
\]
Hence Taylor's formula with integral remainder gives
\[
 \lambda_1(\Sigma+E)=\lambda+L(E)+Q_\lambda(E)+O(\|E\|_{\op}^3).
\]

\paragraph{Calculation of the leading bias.}
Substituting \eqref{eq:cca-first-order-coordinates} into \eqref{eq:cca-correlation-quadratic} gives
\begin{equation}\label{eq:lambda-second-root}
\begin{aligned}
 Q_\lambda(E)={}&-\tfrac12L(E)(\epsilon_X+\epsilon_Y)
                      +\tfrac\lambda8(\epsilon_X-\epsilon_Y)^2\\
 &+\sum_{\ell=2}^m
  \frac{\lambda(\zeta_{X,\ell}^2+\zeta_{Y,\ell}^2)
                  +2\lambda_\ell\zeta_{X,\ell}\zeta_{Y,\ell}}
       {2(\lambda^2-\lambda_\ell^2)}
 +\frac1{2\lambda}\left(\sum_{\ell=m+1}^p\zeta_{X,\ell}^2
                         +\sum_{\ell=m+1}^q\zeta_{Y,\ell}^2\right).
\end{aligned}
\end{equation}
Although the Taylor expansion requires a small perturbation, the quadratic term $Q_\lambda(E)$ is well-defined for arbitrary $E$. We evaluate it at $\Psi=ZZ^\top-\Sigma$ to compute its expectation, using $\E Q_\lambda(\Delta)=n^{-1}\E Q_\lambda(\Psi)$. Writing $\xi=\xi_1$ and $\eta=\eta_1$, Lemma~\ref{lem:gaussian-covariance-increments} gives
\begin{equation}\label{eq:root-bias-moments}
\begin{aligned}
 \E\{L(\Psi)\epsilon_X(\Psi)\}
 &=\E\{L(\Psi)\epsilon_Y(\Psi)\}=\lambda(1-\lambda^2),\\
 \E\{\epsilon_X(\Psi)-\epsilon_Y(\Psi)\}^2&=4(1-\lambda^2),\\
 \E\zeta_{X,\ell}(\Psi)^2=\E\zeta_{Y,\ell}(\Psi)^2&=1-\lambda^2,\\
 \E\{\zeta_{X,\ell}(\Psi)\zeta_{Y,\ell}(\Psi)\}
 &=-\lambda\lambda_\ell(1-\lambda^2),\qquad 2\le\ell\le m.
\end{aligned}
\end{equation}
We show two representative substitutions. In the leading block,
\eqref{eq:gaussian-covariance-increments} gives
\[
 \E\{(\xi\eta-\lambda)(\xi^2-1)\}=2\lambda,\qquad
 \E(\xi^2-1)^2=2,\qquad
 \E\{(\xi^2-1)(\eta^2-1)\}=2\lambda^2.
\]
Since
$L(\Psi)=(\xi\eta-\lambda)-\lambda\{(\xi^2-1)+(\eta^2-1)\}/2$,
these identities yield
\[
 \E\{L(\Psi)\epsilon_X(\Psi)\}=\lambda(1-\lambda^2),\qquad
 \E\{\epsilon_X(\Psi)-\epsilon_Y(\Psi)\}^2=4(1-\lambda^2),
\]
and exchange of the views gives the $Y$-side identity. For $2\le\ell\le m$, $\zeta_{X,\ell}(\Psi)=\xi_\ell(\eta_1-\lambda\xi_1)$ and $\zeta_{Y,\ell}(\Psi)=\eta_\ell(\xi_1-\lambda\eta_1)$. Applying \eqref{eq:gaussian-covariance-increments} to the four centered products in these linear combinations gives
\[
\begin{aligned}
 \E\zeta_{X,\ell}^2
 &=1+\lambda^2-2\lambda^2=1-\lambda^2,\\
 \E(\zeta_{X,\ell}\zeta_{Y,\ell})
 &=\lambda\lambda_\ell-\lambda\lambda_\ell
   -\lambda\lambda_\ell+\lambda^3\lambda_\ell
   =-\lambda\lambda_\ell(1-\lambda^2).
\end{aligned}
\]
The same calculation gives $\E\zeta_{Y,\ell}^2=1-\lambda^2$. For $\ell>m$, each additional coordinate has the same second moment, with no cross term in \eqref{eq:lambda-second-root}. These calculations establish the last two identities in \eqref{eq:root-bias-moments}. The contribution of the leading canonical pair to $\E Q_\lambda(\Psi)$ is $-\lambda(1-\lambda^2)/2$. For each $2\le\ell\le m$, the contribution is
\[
 \frac{2\lambda(1-\lambda^2)
       -2\lambda\lambda_\ell^2(1-\lambda^2)}
      {2(\lambda^2-\lambda_\ell^2)}
 =\frac{\lambda(1-\lambda^2)(1-\lambda_\ell^2)}
       {\lambda^2-\lambda_\ell^2}.
\]
and each extra coordinate contributes $(1-\lambda^2)/(2\lambda)$. Summing these terms yields
\begin{equation}\label{eq:root-quadratic-expectation}
 \E Q_\lambda(\Psi)=\frac{1-\lambda^2}{2}
 \left\{-\lambda+2\lambda\sum_{\ell=2}^m
              \frac{1-\lambda_\ell^2}{\lambda^2-\lambda_\ell^2}
                      +\frac{p+q-2m}{\lambda}\right\}.
\end{equation}
The paired augmented functional is
\begin{equation}\label{eq:paired-lambda-functional}
\begin{aligned}
 F_\lambda(E)={}&\lambda_1(\Sigma+E)
 -\gamma_X k_X\{\lambda_1(\Sigma+E)\}
                          u(\Sigma+E)^\top E_Xu(\Sigma+E)\\
 &-\gamma_Y k_Y\{\lambda_1(\Sigma+E)\}
                          v(\Sigma+E)^\top E_Yv(\Sigma+E).
\end{aligned}
\end{equation}
Its quadratic coefficient is $H_\lambda=Q_\lambda-B_{\rm marg}$, where
\begin{equation}\label{eq:marginal-root-quadratic}
\begin{aligned}
 B_{\rm marg}(E)={}&\gamma_X\{k_X'(\lambda)L(E)\epsilon_X
                       +2k_X(\lambda)h_X(E)^\top E_Xe\}\\
 &+\gamma_Y\{k_Y'(\lambda)L(E)\epsilon_Y
                       +2k_Y(\lambda)h_Y(E)^\top E_Yf\}.
\end{aligned}
\end{equation}
For $2\le\ell\le m$, another direct application of
Lemma~\ref{lem:gaussian-covariance-increments} gives
\[
\begin{aligned}
 \E\{\zeta_{X,\ell}(\Psi)\Psi_{X,\ell1}\}&=0,&
 \E\{\zeta_{Y,\ell}(\Psi)\Psi_{X,\ell1}\}&=\lambda_\ell(1-\lambda^2),\\
 \E\{\zeta_{Y,\ell}(\Psi)\Psi_{Y,\ell1}\}&=0,&
 \E\{\zeta_{X,\ell}(\Psi)\Psi_{Y,\ell1}\}&=\lambda_\ell(1-\lambda^2).
\end{aligned}
\]
For example,
\[
\begin{aligned}
 \E\{\zeta_{X,\ell}(\Psi)\Psi_{X,\ell1}\}
 &=\lambda-\lambda=0,\\
 \E\{\zeta_{Y,\ell}(\Psi)\Psi_{X,\ell1}\}
 &=\lambda_\ell-\lambda^2\lambda_\ell
   =\lambda_\ell(1-\lambda^2),
\end{aligned}
\]
where each covariance on the right is obtained from
\eqref{eq:gaussian-covariance-increments}. The $Y$ formulas follow by
exchanging the two views.
The leading contribution to $\E(h_X^\top\Psi_Xe)$ is
$-\E\epsilon_X^2/2=-1$. By
\eqref{eq:cca-first-order-coordinates}, a paired non-leading index
contributes
\[
 \frac{\lambda\E(\zeta_{X,\ell}\Psi_{X,\ell1})
            +\lambda_\ell\E(\zeta_{Y,\ell}\Psi_{X,\ell1})}
                {\lambda^2-\lambda_\ell^2}
 =\frac{\lambda_\ell^2(1-\lambda^2)}{\lambda^2-\lambda_\ell^2}.
\]
Extra coordinates contribute zero. Hence
\begin{equation}\label{eq:marginal-rotation-moments}
 \E\{h_X(\Psi)^\top\Psi_Xe\}
 =\E\{h_Y(\Psi)^\top\Psi_Yf\}
 =-1+(1-\lambda^2)\sum_{\ell=2}^m
                    \frac{\lambda_\ell^2}{\lambda^2-\lambda_\ell^2}.
\end{equation}
Using $\kappa=\gamma_Xk_X+\gamma_Yk_Y$ gives
\begin{equation}\label{eq:marginal-root-expectation}
 \E B_{\rm marg}(\Psi)=\lambda(1-\lambda^2)\kappa'(\lambda)
 +2\kappa(\lambda)\left\{-1+(1-\lambda^2)\sum_{\ell=2}^m
                         \frac{\lambda_\ell^2}{\lambda^2-\lambda_\ell^2}\right\}.
\end{equation}
Subtracting from \eqref{eq:root-quadratic-expectation} yields
\begin{equation}\label{eq:lambda-bias-explicit}
\begin{aligned}
 \E H_\lambda(\Psi)={}&\frac{1-\lambda^2}{2}
          \left\{\frac{p+q-2m}{\lambda}-\lambda\right\}
   +2\kappa(\lambda)-\lambda(1-\lambda^2)\kappa'(\lambda)\\
 &+(1-\lambda^2)\sum_{\ell=2}^m
   \frac{\lambda(1-\lambda_\ell^2)-2\kappa(\lambda)\lambda_\ell^2}
        {\lambda^2-\lambda_\ell^2}.
\end{aligned}
\end{equation}
Thus $\mathcal C_{\lambda,n}=n^{-1}\E H_\lambda(\Psi)$, and replacing
the correlations by their sample values gives exactly
\eqref{eq:C-lambda-hat}. With no unpaired observations and $p=q=1$,
this is $-\widehat\lambda_1(1-\widehat\lambda_1^2)/(2n)$.

\paragraph{Controlling the remainder bias.}
The compact formulas give
\[
\begin{aligned}
 |Q_\lambda(E)|
 &\le\tfrac12(\|\zeta_X\|\|h_X\|+\|\zeta_Y\|\|h_Y\|)
          +\tfrac14|L|(|\epsilon_X|+|\epsilon_Y|)
 \le C\|E\|_{\op}^2,\\
 |B_{\rm marg}(E)|
 &\le C\{|L|(|\epsilon_X|+|\epsilon_Y|)
             +\|E\|_{\op}(\|h_X\|+\|h_Y\|)\}
 \le C\|E\|_{\op}^2.
\end{aligned}
\]
Hence $|H_\lambda(E)|\le C\|E\|_{\op}^2$ globally as a polynomial,
without summing dimension-dependent coordinate bounds.
For the augmentation along the same path, define
\[
 q_X(t)=k_X\{\lambda(t)\}u(t)^\top E_Xu(t).
\]
At $t=0$,
\[
 q_X(0)=k_X(\lambda)\epsilon_X,\qquad
 q_X'(0)=k_X'(\lambda)L(E)\epsilon_X
             +2k_X(\lambda)h_X(E)^\top E_Xe.
\]
The first two derivatives of $k_X$ are uniformly bounded on the local
neighborhood. By the chain and product rules together with
Lemma~\ref{lem:local-expansions},
\[
 \sup_{0\le t\le1}|q_X''(t)|\le C\|E\|_{\op}^3.
\]

Therefore,
\[
 t q_X(t)=t k_X(\lambda)\epsilon_X
 +t^2\{k_X'(\lambda)L(E)\epsilon_X
             +2k_X(\lambda)h_X(E)^\top E_Xe\}
 +O(t^3\|E\|_{\op}^3).
\]
The corresponding expansion for the $Y$-side correction follows by exchanging the two views. Combining these expansions with the Taylor expansion of
$\lambda_1$ gives
\begin{equation}\label{eq:lambda-paired-taylor}
 F_\lambda(E)-F_\lambda(0)-DF_\lambda(0)[E]
 =H_\lambda(E)+O(\|E\|_{\op}^3).
\end{equation}

Lemma~\ref{lem:local-expansions} and the bounded derivatives of $k_X,k_Y$ verify the regularity conditions of Lemma~\ref{lem:augmented-expansion}. Choose $\delta>0$ sufficiently small so that $\{S:\|S-\Sigma\|_{\op}<2\delta\}$ lies in this neighborhood and $\delta<\lambda_{\min}(\Sigma)$. Applying Lemma~\ref{lem:augmented-expansion} with the quadratic approximation $H_\lambda$ yields
\begin{equation}\label{eq:lambda-conditional-mean-variance}
 \mu_{\lambda,n}:=
 \E\{R_{F_\lambda}(\Delta)\mid\mathcal E_n\}
 =\mathcal C_{\lambda,n}+O((d/n)^{3/2}).
\end{equation}
The remainder term also accounts for the difference between the conditional and unconditional expectations of $H_\lambda(\Delta)$. Thus, the second-order bias is estimated by $\widehat{\mathcal C}_{\lambda,n}$ in \eqref{eq:C-lambda-hat}.

\subsection{Proof of Theorem~\ref{thm:lambda-ci}}
\begin{proof}
As in the preceding derivation, we work in the population canonical coordinates defined in \eqref{eq:fixed-canonical-transform}. Apply Lemma~\ref{lem:augmented-expansion} with
\[
 \theta(S)=\lambda_1(S),\qquad
 B_X(S)=B_{\lambda,X}(S),\qquad
 B_Y(S)=B_{\lambda,Y}(S),\qquad
 \widehat{\mathcal C}_n=\widehat{\mathcal C}_{\lambda,n}.
\]
Lemma~\ref{lem:local-expansions} verifies the required regularity conditions, and \eqref{eq:generic-augmented-estimator} coincides with \eqref{eq:lambda-ss-estimator} on $\mathcal E_n$. The corresponding influence function and augmentation functions are $g_\lambda,a_\lambda,b_\lambda$ from Section~\ref{sec:lambda-influence-proof}. Thus, the linear term in \eqref{eq:generic-linear-average} is
\begin{equation}\label{eq:lambda-linear-averages}
\begin{aligned}
 \mathcal L_{\lambda,n}
 ={}&\frac1n\sum_{i=1}^n
 \{g_\lambda(X_i,Y_i)
   -\gamma_Xa_\lambda(X_i)
   -\gamma_Yb_\lambda(Y_i)\}\\
 &+\frac1{n+N_X}\sum_{r=1}^{N_X}a_\lambda(X_r^U)
 +\frac1{n+N_Y}\sum_{s=1}^{N_Y}b_\lambda(Y_s^U).
\end{aligned}
\end{equation}

Lemma~\ref{lem:augmented-expansion} and \eqref{eq:lambda-conditional-mean-variance} give
\begin{equation}\label{eq:lambda-before-stochastic-centering}
\begin{aligned}
 \widehat\lambda_1^{\mathrm{Semi}}-\lambda_1
 &=\mathcal L_{\lambda,n}+\mu_{\lambda,n}
   -\widehat{\mathcal C}_{\lambda,n}
   +O_p(\sqrt d/n),\\
 \mu_{\lambda,n}
 &=\mathcal C_{\lambda,n}+O((d/n)^{3/2}).
\end{aligned}
\end{equation}

For covariance matrices $S$ in a sufficiently small neighborhood of $\Sigma$, the leading canonical correlation is bounded away from zero and one, and the leading eigengap and $1-\gamma_X\gamma_Y\lambda_1^4$ are bounded away from zero. Moreover, $\kappa$ and its first two derivatives are uniformly bounded. Each summand in \eqref{eq:lambda-bias-explicit} is uniformly Lipschitz in $\lambda_1$ and $\lambda_\ell^2$, while the term involving $p+q-2m$ has Lipschitz constant at most $Cd$. Therefore,
\begin{equation}\label{eq:bias-plugin-rate}
 |\widehat{\mathcal C}_{\lambda,n}-\mathcal C_{\lambda,n}|
 \le\frac Cn\left\{
 d|\widehat\lambda_1-\lambda_1|
 +\sum_{\ell=2}^m
 |\widehat\lambda_\ell^2-\lambda_\ell^2|
 \right\}
 \le\frac{Cd}{n}\|\Delta\|_{\op}
 =O_p(d^{3/2}/n^{3/2}),
\end{equation}
by \eqref{eq:all-correlations-lipschitz}.
Substituting these bounds into
\eqref{eq:lambda-before-stochastic-centering} yields
\begin{equation}\label{eq:lambda-asymptotic-linear}
 \widehat\lambda_1^{\mathrm{Semi}}-\lambda_1
 =\mathcal L_{\lambda,n}
 +O_p\left(\frac{\sqrt d}{n}
          +\frac{d^{3/2}}{n^{3/2}}\right).
\end{equation}
Under $d=o(n^{2/3})$, the remainder is $o_p(n^{-1/2})$.

By independence of the paired and unpaired samples, $\Var(\sqrt n\,\mathcal L_{\lambda,n})=\sigma_{\lambda,n}^2$. The bounded quadratic coefficients, \eqref{eq:gaussian-fourth-moment}, \eqref{eq:common-lyapunov-bound}, and \eqref{eq:sigma-lambda-bounds} verify the conditions of Lemma~\ref{lem:prelim-lyapunov} with $\eta=2$. The function $s_\gamma^2(\lambda)$ is uniformly Lipschitz in a neighborhood of $\lambda_1$. Hence, \eqref{eq:direct-lipschitz} gives
\[
\widehat\sigma_\lambda^2-\sigma_{\lambda,n}^2
=O_p(\sqrt{d/n}).
\]
Together with the positive lower bound in \eqref{eq:sigma-lambda-bounds}, this implies $\widehat\sigma_\lambda^2/\sigma_{\lambda,n}^2 \xrightarrow{p}1$. Equation~\eqref{eq:prelim-studentization} then establishes the limiting distribution. The estimator and variance estimate are invariant to simultaneous sign changes of the leading canonical directions. Since $\Pr(\mathcal E_n)\to1$, their definitions outside $\mathcal E_n$ do not affect the limiting distribution. Consequently, for any fixed $\alpha\in(0,1)$, the confidence interval
\begin{equation}\label{eq:lambda-ci}
[\widehat\lambda_1^{\mathrm{Semi}}
\pm z_{1-\alpha/2}\widehat\sigma_\lambda/\sqrt n]
\end{equation}
has asymptotic coverage $1-\alpha$.
\end{proof}

\setcounter{table}{0}
\renewcommand{\thetable}{S\arabic{table}}
\renewcommand{\theHtable}{supp.\arabic{table}}
\raggedbottom
\section{Additional numerical results}

\subsection{Variance reduction}
\label{sec:additional-mechanism}
We use the Gaussian model in Section~\ref{sec:rank-two-comparison} at $n=3000$, $p=q=26$, and $N_X=N_Y=2000$. We retain the same AR$(0.3)$ marginal covariances and canonical directions, with 13 nonzero coefficients in each leading direction. We fix $\lambda_1=0.85$ and vary $\lambda_2\in\{0,0.25,0.50,0.75\}$, with all other correlations zero. In both views, coordinate 26 is zero in the leading direction and nonzero in the second. Coordinate 14 is zero in both directions. PW and SW use second moments about the known zero mean and the same paired samples within each replication. Table~\ref{tab:review-mechanism} compares the population asymptotic variance reduction from with the empirical reduction
\[
 100\left\{1-
 \frac{\operatorname{Var}_r(\widehat u^{\mathrm{Semi, (r)}}_{1j})}
      {\operatorname{Var}_r(\widehat u^{\mathrm{(r)}}_{1j})}\right\}.
\]
Here $\widehat u^{\mathrm{(r)}}_{1j}$ is the coordinate estimate in replication $r$, and $\operatorname{Var}_r$ denotes variance across replications.
\begin{table}[!ht]\centering
\setlength{\tabcolsep}{6pt}
\caption{Null-coordinate variance reductions (\%) relative to PW at $n=3000$, $p=q=26$, and $N_X=N_Y=2000$, averaged over 500 replications. The population reduction is identical for $X$ and $Y$. Negative values indicate larger empirical SW variance.}
\label{tab:review-mechanism}
\begin{tabular}{@{}rrrrr@{}}
\toprule
$\lambda_2$ & Coordinate & Population & Empirical $X$ & Empirical $Y$\\
\midrule
0.00 & 26 & 0.00 & $-0.24$ & $-0.16$\\
0.25 & 26 & 0.37 & $0.56$ & $0.19$\\
0.50 & 26 & 2.43 & $3.34$ & $2.89$\\
0.75 & 26 & 12.47 & $16.04$ & $15.96$\\
\addlinespace[2pt]
0.75 & 14 & 0.00 & $-0.44$ & $-0.75$\\
\bottomrule
\end{tabular}
\end{table}

At coordinate 26, the predicted variance reduction increases from $0\%$ when $\lambda_2=0$ to $12.47\%$ when $\lambda_2=0.75$. At the latter setting, empirical reductions are about $16\%$ for both views. At coordinate 14, the predicted gain is zero even at $\lambda_2=0.75$, and empirical changes are less than one percentage point in magnitude. The results agree with Proposition~\ref{prop:var-reduction}: coordinate 26 exhibits variance reduction when $\lambda_2>0$, while coordinate 14 shows little change. The distinction arises because the second canonical direction has a nonzero coefficient at coordinate 26 but a zero coefficient at coordinate 14.

\subsection{Coverage and length of confidence intervals for the leading canonical correlation}
\label{sec:additional-correlation}
\begin{table}[!ht]\centering
\setlength{\tabcolsep}{10pt}
\caption{Coverage (\%) and mean length of 95\% confidence intervals
for $\lambda_1=0.85$ from 500 replications. Each method uses its
bias-corrected estimate as both the interval center and the variance
plug-in. SW$_X$ and SW$_{XY}$ use $(2000,0)$ and $(2000,2000)$
unpaired observations, respectively.}
\label{tab:review-correlation}
\begin{tabular}{@{}rrlrr@{}}
\toprule
$n$ & $p=q$ & Method & Coverage & Mean length\\
\midrule
2000 & 18 & PW & 95.2 & 0.02435\\
2000 & 18 & SW$_X$ & 95.6 & 0.02204\\
2000 & 18 & SW$_{XY}$ & 95.8 & 0.02087\\
\addlinespace[2pt]
3000 & 22 & PW & 96.2 & 0.01988\\
3000 & 22 & SW$_X$ & 95.6 & 0.01839\\
3000 & 22 & SW$_{XY}$ & 96.6 & 0.01753\\
\addlinespace[2pt]
4000 & 26 & PW & 96.6 & 0.01718\\
4000 & 26 & SW$_X$ & 96.6 & 0.01612\\
4000 & 26 & SW$_{XY}$ & 96.8 & 0.01543\\
\bottomrule
\end{tabular}
\end{table}

We compare 95\% confidence intervals for $\lambda_1=0.85$ under the same setting in Gaussian rank-two model in Section~\ref{sec:rank-two-comparison}. We use $(n,p,q)=(2000,18,18)$, $(3000,22,22)$, and $(4000,26,26)$. PW uses $(N_X,N_Y)=(0,0)$, SW$_X$ uses $(2000,0)$, and SW$_{XY}$ uses $(2000,2000)$. At each setting, all methods share the same 500 paired datasets, and the two semi-paired methods share the same $X$-only samples. For all methods, we use the bias-corrected estimator in \eqref{eq:lambda-ss-estimator} as the interval center, with PW is its special case with no unpaired observations. We estimate the variance using \eqref{eq:sigma-lambda-hat}
\[
 \widehat\sigma_\lambda^2
 =(1-\widehat\lambda_1^2)^2
 \frac{1-\tfrac12(\gamma_X+\gamma_Y)\widehat\lambda_1^2}
      {1-\gamma_X\gamma_Y\widehat\lambda_1^4},
 \qquad \gamma_X=\frac{N_X}{n+N_X},\quad
 \gamma_Y=\frac{N_Y}{n+N_Y}.
\]
The variance formula uses the corresponding canonical correlation estimates. We report coverage and mean length for the intervals $[\widehat\lambda_1^{\mathrm{Semi}}\pm z_{0.975}\widehat\sigma_\lambda/\sqrt n]$. Coverage is the fraction containing $\lambda_1=0.85$, and mean length is the average interval width over 500 replications. 

As shown in Table~\ref{tab:review-correlation}, coverage ranges from $95.2\%$ to $96.8\%$, close to the nominal $95\%$ level. The main benefit of auxiliary data is shorter intervals with comparable coverage: relative to PW, SW$_X$ reduces mean length by $6.2$--$9.5\%$ and SW$_{XY}$ by $10.2$--$14.3\%$. The relative length reduction decreases as $n$ increases because $N_X$ and $N_Y$ remain fixed, reducing the auxiliary fractions $\gamma_X$ and $\gamma_Y$.

\end{document}